\documentclass[
apjs,
twocolappendix,
twocolumn
]{aastex631}
\submitjournal{ApJS}

\usepackage[T1]{fontenc}
\usepackage{CJKutf8}
\usepackage{hyperref}
\usepackage{amsmath} 
\usepackage{physics}
\usepackage{amssymb}
\usepackage{enumerate}
\usepackage{bm}

\shorttitle{
Multi-frequency general-relativistic radiation transport in \texttt{Gmunu}
}
\shortauthors{Cheong et al.}
\graphicspath{{./}{figures/}}

\begin{document}

\title{General-relativistic radiation transport scheme in \texttt{Gmunu} \\
\textbf{I}: Implementation of two-moment based multi-frequency radiative transfer and code tests
}

\author[0000-0003-1449-3363]{Patrick Chi-Kit Cheong \begin{CJK*}{UTF8}{bkai}(張志杰)\end{CJK*}}
\email{patrick.cheong@berkeley.edu}
\affiliation{Department of Physics \& Astronomy, University of New Hampshire, 9 Library Way, Durham NH 03824, USA}
\affiliation{Department of Physics, University of California, Berkeley, Berkeley, CA 94720, USA}

\author[0000-0003-3453-7394]{Harry Ho-Yin Ng}
\affiliation{Institut f\"ur Theoretische Physik, Goethe Universit\"at, Max-von-Laue-Str. 1, 60438 Frankfurt am Main, Germany}

\author[0000-0002-1307-1401]{Alan Tsz-Lok Lam}
\affiliation{Max Planck Institute for Gravitational Physics (Albert Einstein Institute), Am Mühlenberg 1, Postdam-Golm 14476, Germany}

\author[0000-0003-4297-7365]{Tjonnie Guang Feng Li}
\affiliation{Department of Physics and Astronomy, KU Leuven, Celestijnenlaan 200D, B-3001 Leuven, Belgium}
\affiliation{Department of Electrical Engineering (ESAT), KU Leuven, Kasteelpark Arenberg 10, B-3001 Leuven, Belgium }
\affiliation{Department of Physics, The Chinese University of Hong Kong, Shatin, N.T., Hong Kong}



\begin{abstract}
	We present the implementation of two-moment based general-relativistic multi-group radiation transport module in the \texttt{G}eneral-relativistic \texttt{mu}ltigrid \texttt{nu}merical (\texttt{Gmunu}) code.
	On top of solving the general-relativistic magnetohydrodynamics and the Einstein equations with conformally flat approximations, the code solves the evolution equations of the zeroth- and first-order moments of the radiations {in Eulerian frame}.
	Analytic closure relation is used to obtain the higher order moments and close the system.
	The finite-volume discretisation has been adopted for the radiation moments.
	The advection in spatial and frequency spaces are handled explicitly.
	In addition, the radiation-matter interaction terms, which are very stiff in the optically thick region, are solved implicitly.
	Implicit-explicit Runge-Kutta schemes are adopted for time integration.
	We test the implementation with a number of numerical benchmarks from frequency-integrated to frequency dependent cases.
	Furthermore, we also illustrate the astrophysical applications in hot neutron star and core-collapse supernovae modellings, and compare with other neutrino transport codes.
\end{abstract}



\section{\label{sec:intro}Introduction}
Radiation transport plays a crucial role in many high energy astrophysical events.
For instance, radiation cooling and transport can significantly affect the structure and behaviour of black hole accretion disks (see, e.g., \cite{
2012ApJS..201....9F,
2013MNRAS.429.3533S,
2014ApJ...796...22F,
2014MNRAS.441.3177M,
2016ApJ...826...23T,
2018ApJ...857....1F}).
On the other hand, neutrinos are responsible for the transport of energy and lepton number in dense and hot scenarios. 
For example, neutrinos largely determine the properties of the matter ejected by neutron star mergers.
This matter is responsible for part of the observational electromagnetic signatures powered by nuclear reactions, as well as the contribution to astrophysical nucleosynthesis (see, e.g. \cite{
2009ApJ...690.1681D, 
2014MNRAS.441.3444M, 
2014MNRAS.443.3134P, 
2015PhRvD..91f4059S, 
2015ASPC..498..121R, 
2016PhRvD..93d4019F, 
2016PhRvD..93l4046S, 
2017ApJ...850L..37P, 
2017ApJ...846..114F, 
2018ApJ...860...64F, 
2019MNRAS.482.3373F, 
2019ApJ...886L..30N, 
2019PhRvD.100b3008M, 
2020PhRvD.101h3029F, 
2020ApJ...901..122F, 
2021PhRvL.126p2701E, 
2022MNRAS.509.1377J}).
Not only in the context of neutron star mergers, neutrinos also significantly affect the dynamics of the core-collapse of massive stars, and are responsible for powering the explosion as supernovae (see, e.g., \cite{
2012ARNPS..62..407J,
2013RvMP...85..245B,
2015PASA...32....9F,
2015ApJ...807L..31L, 
2015ApJ...808L..42M, 
2018ApJ...865...81O, 
2020MNRAS.491.2715B, 
2021ApJ...915...28B, 
2021Natur.589...29B}).
In order to have a better understanding of such high energy astrophysical systems, not only do we need general relativistic magnetohydrodynamics simulations, but a proper treatment for radiation transport is essential.

The full Boltzmann transport equation needs to be solved for the evolution of radiation fields.
Although in principle this can be solved exactly by using
the short characteristic method \citep{2012ApJS..199....9D}, 
the $S_N$ schemes \citep{2012ApJS..199...17S, 2014ApJS..214...16N, 2017ApJS..229...42N, 2020MNRAS.496.2000C, 2023arXiv230204283W}, 
the spherical harmonics schemes \citep{2010JCoPh.229.5597M, 2013JCoPh.242..648R}, 
the lattice Boltzmann methods \citep{2020MNRAS.498.3374W}, 
the method of characteristic moment closure \citep{2020ApJ...891..118R} and 
the Monte Carlo method \citep{2012ApJ...755..111A, 2019PhRvD.100b3008M, 2021ApJ...920...82F, 2022arXiv220912472K}, 
solving the Boltzmann equation exactly is usually not practical due to the high computational cost.
In practice, simplified versions of the Boltzmann transport equation are solved approximately.

One widely adopted approach is to solve only the first few moments of the radiation distribution function based on the truncated moment formalism \citep{1981MNRAS.194..439T, 2011PThPh.125.1255S, 2013PhRvD..87j3004C}.
For example, the flux-limited diffusion approximation scheme solves only the zeroth moment \citep{1981ApJ...248..321L, 1981JQSRT..26..385P}.
In this scheme, only the information of radiation intensity is available, while the propagation directions are not.
The direction of radiation flow can be retained in an averaged fashion by solving the first moment as well; this is known as the two-moment scheme \citep{1984JQSRT..31..149L, 1999CRASM.329..915D}.
This approach has been applied in the context of neutron star merger \citep{
2014ApJ...789L..39W,
2015PhRvD..91l4021F, 
2016PhRvD..94l3016F, 
2016PhRvD..93d4019F, 
2015PhRvD..91f4059S, 
2022MNRAS.512.1499R} (see also the review \cite{2022arXiv220902538F}), 
core-collapse supernovae \citep{
2015ApJS..219...24O, 
2015MNRAS.453.3386J, 
2016ApJ...831...98R,
2016ApJS..222...20K, 
2018ApJ...865...81O, 
2019ApJS..241....7S, 
2021ApJS..253...52L,
2023arXiv230212089S} (see also the review \cite{2020LRCA....6....4M}),
black hole accretions \citep{
2011MNRAS.417.2899Z,
2012ApJS..201....9F,
2013MNRAS.429.3533S,
2014ApJ...796...22F,
2014MNRAS.441.3177M,
2016ApJ...826...23T,
2018ApJ...857....1F},
and other purposes \citep{
2007A&A...464..429G, 
2011A&A...529A..35C, 
2013ApJS..206...21S,
2013ApJ...772..127T,
2019ApJS..241...28R, 
2019ApJS..242...20M,
2020MNRAS.495.2285W,
2020ApJ...900...71A,
2023CoPhC.28408630L} as well.
Despite the recent progress of radiation transport modelling, the implementation of general-relativistic multi-species multi-group radiation magnetohydrodynamics codes which include fully coupled radiation-matter interactions are still not very common (but see an notable example \cite{2016ApJS..222...20K}), but are essential for astrophysics modelling.

In this work, we extend \texttt{Gmunu} \citep{2020CQGra..37n5015C, 2021MNRAS.508.2279C, 2022ApJS..261...22C, 2023_leakage} by implementing the two-moment based multi-frequency multi-species general-relativistic radiation transport module.
In particular, we evolve the zeroth- and first-order moments, and adopt the maximum-entropy closure \citep{1978JQSRT..20..541M} to close the system.
The advection in spatial space is handled by the standard high-resolution-shock-capturing method with a small modification on the Harten, Lax and van Leer (HLL) flux \citep{harten1983upstream} in order to reduce the asymptotic diffusion limit in the high opacity region.
The advection in frequency space is evolved explicitly in a way that the energy-momentum is conserved \citep{2010ApJS..189..104M, 2016ApJS..222...20K}. 
As such, our code is able to capture the Doppler and gravitational redshift effects. 
The radiation-matter interaction terms are solved implicitly since they are very stiff in the optically thick region.
As in our previous work \cite{2022ApJS..261...22C}, we adopt the Implicit-explicit (IMEX) Runge-Kutta time integrators (see, e.g., \cite{ASCHER1997151, pareschi2005implicit}) to implicitly handle the stiff source terms while keeping the time step reasonable.
These schemes have been applied and tested in several radiation hydrodynamics codes, e.g. \cite{2020MNRAS.495.2285W, 2020ApJ...900...71A, 2021ApJS..253...52L, 2022arXiv221100027I}.
We test the implementation with a number of numerical benchmarks, ranging from special-relativistic to general-relativistic, from optically thick to optically thin and from frequency-integrated to frequency-dependent cases.
Moreover, we also include core-collapse supernova and hot neutron star modelling as astrophysical application examples.
We then compare the result with other neutrino (magneto)hydrodynamics codes.

Accompanying this work, we have also developed a new neutrino microphysics library \texttt{Weakhub} \citep{2023_weakhub}.
This library includes the state-of-the-art neutrino microphysics, and provides advanced neutrino opacities and kernels that are essential to neutron star mergers and core-collapse supernovae modellings.
Since the main focus of this work is to present and test the implementation of our new radiative transfer hydrodynamics module, the details of the neutrino microphysics are not included in this paper.
For a formal discussion of neutrino microphysics, we refer readers to \cite{2023_weakhub}.

The paper is organised as follows.
In section~\ref{sec:formulations} we outline the formalism we used in this work.
The details of the methodology and implementation of our radiation transport module are presented in section~\ref{sec:numerical_methods}.
The code tests and results with idealised neutrino opacities are presented in section~\ref{sec:numerical_tests}.
The comparison of different neutrino transport code with a conventional set of neutrino opacities in the context of core-collapse supernovae and hot neutron star are presented in section~\ref{sec:application_tests}.
This paper ends with a discussion in section~\ref{sec:conclusions}.

{Unless} explicitly stated, the unit of which the speed of light $c$, gravitational constant $G$, solar mass $\rm{M_{\odot}}$ and the Boltzmann constant $k_{\rm{B}}$ are all {equal} to one ($c=G={\rm M_{\odot}}=k_{\text{B}}= 1$).
Greek indices, running from 0 to 3, are used for 4-quantities while the Roman indices, running from 1 to 3, are used for 3-quantities.

\section{Formulations}\label{sec:formulations} 
The comoving-frame zeroth-, first-, second- and third-order moments are defined as \citep{2013PhRvD..87j3004C, 2020LRCA....6....4M}
\begin{equation}
	\begin{aligned}
		\mathcal{J}\left( x^\mu, \nu \right) &\equiv \frac{\nu}{4 \pi} \int f \left( x^\mu, \nu, \Omega \right) \dd{\Omega} ,\\
		\mathcal{H}^\alpha \left( x^\mu, \nu \right) &\equiv \frac{\nu}{4 \pi} \int \ell^\alpha f \left( x^\mu, \nu, \Omega \right) \dd{\Omega} ,\\
		\mathcal{K}^{\alpha\beta} \left( x^\mu, \nu \right) &\equiv \frac{\nu}{4 \pi} \int \ell^\alpha \ell^\beta f \left( x^\mu, \nu, \Omega \right) \dd{\Omega} , \\
		\mathcal{L}^{\alpha\beta\gamma} \left( x^\mu, \nu \right) &\equiv \frac{\nu}{4 \pi} \int \ell^\alpha \ell^\beta \ell^\gamma f \left( x^\mu, \nu, \Omega \right) \dd{\Omega} , 
	\end{aligned}
\end{equation}
where $f$ is the distribution function, $\ell^\alpha$ is the unit three-vector tangent to the three-momentum in the comoving frame, namely $u_\mu \ell^\mu = 0$.
$\nu$ is the frequency of radiation observed in the comoving frame while $\dd{\Omega}$ is the solid angle in the comoving frame.

The monochromatic energy-momentum tensor $\mathcal{T}^{\mu\nu}$ and the corresponding third-rank momentum moment $\mathcal{U}^{\mu\nu\rho}$ can be Lagrangian decomposed with respect to the comoving observer with four-velocity $u^\mu$ as follows:
\begin{align}
	&\mathcal{T}^{\mu\nu} = \mathcal{J} u^\mu u^\nu + \mathcal{H}^\mu u^\nu + u^\mu \mathcal{H}^\nu + \mathcal{K}^{\mu\nu}, \\
	&
	\begin{aligned}\label{eq:uabc}
		\mathcal{U}^{\mu\nu\rho} = & \nu \Big( \mathcal{J} u^\mu u^\nu u^\rho 
		+ \mathcal{H}^\mu u^\nu u^\rho 
		+ u^\mu \mathcal{H}^\nu u^\rho 
		+ u^\mu u^\nu \mathcal{H}^\rho \\
		& \quad + \mathcal{K}^{\mu\nu} u^\rho
		+ \mathcal{K}^{\nu\rho} u^\mu
		+ \mathcal{K}^{\rho\mu} u^\nu
		+ \mathcal{L}^{\mu\nu\rho} \Big)
		,
	\end{aligned}
\end{align}
where $\mathcal{H}^\mu u_\mu$ and $\mathcal{K}^{\mu\nu} u_\mu=0=\mathcal{K}^{\mu\nu} u_\nu$.
The corresponding frequency-integrated energy-momentum tensor of the radiation can be obtained by
\begin{equation}\label{eq:em_tensor}
	T^{\mu\nu}_{\text{rad}} = \int_0^\infty 4 \pi \nu^2 \mathcal{T}^{\mu\nu} \dd{\nu} = \int_0^\infty \mathcal{T}^{\mu\nu} \dd{V_\nu},
\end{equation}
where we have defined $\dd{V_\nu} \equiv 4 \pi \nu^2 \dd{\nu} $.

Alternatively, the monochromatic energy-momentum tensor $\mathcal{T}^{\mu\nu}$ and the third-rank momentum moment $\mathcal{U}^{\mu\nu\rho}$ can be Eulerian decomposed with respect to the Eulerian observer with four-velocity $n^\mu$ as follows
\begin{equation}
	\begin{aligned}
		\mathcal{T}^{\mu\nu} = & \mathcal{E} n^\mu n^\nu + \mathcal{F}^\mu n^\nu + n^\mu \mathcal{F}^\nu + \mathcal{P}^{\mu\nu}, \\
		\mathcal{U}^{\mu\nu\rho} = & \nu \Big( \mathcal{Z} n^\mu n^\nu n^\rho 
		+ \mathcal{Y}^\mu n^\nu n^\rho 
		+ n^\mu \mathcal{Y}^\nu n^\rho 
		+ n^\mu n^\nu \mathcal{Y}^\rho \\
		& \quad + \mathcal{X}^{\mu\nu} n^\rho
		+ \mathcal{X}^{\nu\rho} n^\mu
		+ \mathcal{X}^{\rho\mu} n^\nu
		+ \mathcal{W}^{\mu\nu\rho} \Big)
		,
	\end{aligned}
\end{equation}
where $\mathcal{F}^\mu n_\mu$ and $\mathcal{P}^{\mu\nu} n_\mu=0=\mathcal{P}^{\mu\nu} n_\nu$.

The evolution equations of the radiation can be obtained by
\begin{equation}\label{eq:evolution}
	\nabla_\nu \mathcal{T}^{\mu\nu} - \frac{1}{\nu^2}\frac{\partial}{\partial \nu} \left( \nu^2 \mathcal{U}^{\mu\nu\rho} \nabla_\rho u_\nu \right) = \mathcal{S}^\mu_{\rm{rad}},
\end{equation}
where $\mathcal{S}^\mu_{\rm{rad}}$ is the radiation four-force, which describes the interaction between the radiation and the fluid.

The choice of the radiation four-force $\mathcal{S}_{\text{rad}}$ depends on the type of radiation considered.
In general, the radiation four-force contains the coupling between different radiation species at different frequencies group.
By default, the radiation four-force $\mathcal{S}_{\text{rad}}$ contains the emission and absorption source term $\mathcal{S}^\mu_{\text{E/A}}$, and the elastic (iso-energetic) scattering source term $\mathcal{S}^\mu_{\text{ES}}$:
\begin{equation}\label{eq:source_terms}
	\mathcal{S}^\mu_{\text{rad}} = \mathcal{S}^\mu_{\text{E/A}} + \mathcal{S}^\mu_{\text{ES}},
\end{equation}
and neglecting the frequency/species coupling.
Here, the emission and absorption source term $\mathcal{S}^\mu_{\text{E/A}}$ and the elastic (iso-energetic) scattering source term $\mathcal{S}^\mu_{\text{ES}}$ are defined as
\begin{align}
	&\mathcal{S}^\mu_{\text{E/A}} = \left[ \eta - \kappa_a \mathcal{J} \right] u^\mu - \kappa_a \mathcal{H}^\mu \label{eq:e_and_a_source},\\
	&\mathcal{S}^\mu_{\text{ES}} = - \kappa_s \mathcal{H}^\mu \label{eq:iso_scattering_source},
\end{align}
where $\eta $, $\kappa_a $ and $\kappa_s $ are the radiation emissivity, absorption and scattering coefficients, respectively.

It is worth to point out that, \texttt{Gmunu} has been designed to handle more sophisticated radiation four-forces where the frequency/species coupling are allowed (see section~\ref{sec:rank} below).
Since those interactions are application orientated, the discussion is not included in this section.
An example of such complicated radiation four-force in the context of core-collapse supernovae can be found at section~\ref{sec:nu_int} below.

\subsection{General relativistic radiation hydrodynamics in the reference-metric formalism}\label{sec:GRRHD_equations}
As in our previous work \cite{2021MNRAS.508.2279C, 2022ApJS..261...22C, 2023_leakage}, we adopt $3+1$ reference-metric formalism \citep{2014PhRvD..89h4043M, 2020PhRvD.101j4007M, 2020PhRvD.102j4001B}.
In this formalism, the metric can be written as
\begin{equation}
	\begin{aligned}
	ds^2 = & g_{\mu\nu}dx^\mu dx^\nu \\
		= & -\alpha^2  dt^2 + \gamma_{ij} \left( dx^i + \beta^i dt \right) \left( dx^j + \beta^j dt \right)
	\end{aligned}
\end{equation}
where $\alpha$ is the lapse function, $\beta^i$ is the spacelike shift vector and $\gamma_{ij}$ is the spatial metric.
We adopt a conformal decomposition of the spatial metric $\gamma_{ij}$ with the conformal factor $\psi$:
\begin{equation}
	\gamma_{ij} = \psi^4 \bar{\gamma}_{ij},
\end{equation}
where $\bar{\gamma}_{ij}$ is the conformally related metric.
This conformally related metric can be expressed as the sum of a background \emph{time-independent} reference metric $\hat{\gamma}_{ij}$ and deviations $h_{ij}^{\rm dev}$.
In our current implementation, the reference metric $\hat{\gamma}_{ij}$ is the flat spacetime metric of the chosen coordinate system (i.e. either Cartesian, cylindrical or spherical coordinates).
Note that, in conformally flat approximations, the spacetime deviations are vanishing and the reference metric is the conformally related metric (i.e. $\bar{\gamma}_{ij}=\hat{\gamma}_{ij}$).

The evolution equations of the first two moments of radiations for each species of radiation at each frequency group (equation~\eqref{eq:evolution}) can be written as
\begin{align}
	&\begin{aligned}\label{eq:evolution_e}
		\frac{\partial}{\partial t} \left[ \sqrt{{\gamma}/\hat{\gamma}} \mathcal{E} \right] 
		& + \hat{\nabla}_i \left[\sqrt{{\gamma}/\hat{\gamma}} \left(\alpha {\mathcal{F}}^i - \mathcal{E} \beta^i \right)\right] \\
		& - \alpha \sqrt{{\gamma}/\hat{\gamma}} \frac{1}{\nu^2}\frac{\partial}{\partial \nu} \left[ - \nu^2 n_\mu \mathcal{U}^{\mu\nu\rho} \nabla_\rho u_\nu \right] \\
	 	= &\sqrt{{\gamma}/\hat{\gamma}} \left[ - {\mathcal{F}}^j \partial_j \alpha  + {\mathcal{P}}^{ij}K_{ij} \right] \\
		 &- \alpha \sqrt{{\gamma}/\hat{\gamma}} \mathcal{S}_{\rm{rad}}^\mu n_\mu, 
	\end{aligned} \\
	&\begin{aligned}\label{eq:evolution_f}
		\frac{\partial}{\partial t} \left[ \sqrt{{\gamma}/\hat{\gamma}} \mathcal{F}_i \right] 
		& + \hat{\nabla}_i \left[\sqrt{{\gamma}/\hat{\gamma}} \left( \alpha {\mathcal{P}}^i_{\; j} - {\mathcal{F}}_j \beta^i \right)\right] \\
		& - \alpha \sqrt{{\gamma}/\hat{\gamma}} \frac{1}{\nu^2}\frac{\partial}{\partial \nu} \left[ \nu^2 \gamma_{i \mu} \mathcal{U}^{\mu\nu\rho} \nabla_\rho u_\nu \right] \\
	 	= &\sqrt{{\gamma}/\hat{\gamma}} \left[ - \mathcal{E} \partial_i \alpha + {\mathcal{F}}_k \hat{\nabla}_i \beta^k + \frac{1}{2} \alpha {\mathcal{P}}^{jk} \hat{\nabla}_i \gamma_{jk}\right] \\
		& + \alpha \sqrt{{\gamma}/\hat{\gamma}} \mathcal{S}_{\rm{rad}}^\mu \gamma_{i\mu}, 
	\end{aligned}
\end{align}
where the $\hat{\nabla}_i$ here is the covariant derivatives associated with the reference metric $\hat{\gamma}_{ij}$.

As in our previous work \cite{2021MNRAS.508.2279C, 2022ApJS..261...22C, 2023_leakage}, the evolution equations can be expressed as:
\begin{equation}
	\begin{aligned}
		\partial_t \bm{q} + & \frac{1}{\sqrt{\hat{\gamma}}}\partial_j\left[\sqrt{\hat{\gamma}} \bm{f}^j\right] + \frac{1}{\nu^2}\partial_\nu \left[ \nu^2 \bm{f_\nu} \right] \\
		& = \bm{s}_{\text{grav}} + \bm{s}_{\text{geom}} + \bm{s}_{\text{rad}},
	\end{aligned}
\end{equation}
where we denote
\begin{equation}\label{eq:cons}
	\bm{q} = \begin{bmatrix}
           q_{\mathcal{E}} \\
           q_{\mathcal{F}_{j}} \\
        \end{bmatrix}, 
	\bm{f^i} = 
        \begin{bmatrix}
	   \left(f_{\mathcal{E}}\right)^i \\
           \left(f_{\mathcal{F}_{j}}\right)^i \\
        \end{bmatrix} ,
	\bm{f_\nu} = 
        \begin{bmatrix}
	   {f_\nu}_{\mathcal{E}} \\
           {f_\nu}_{\mathcal{F}_{j}} \\
        \end{bmatrix}
	\bm{s} = 
        \begin{bmatrix}
           s_{\mathcal{E}} \\
           s_{\mathcal{F}_{j}} \\
        \end{bmatrix}
	.
\end{equation}
Note that the subscript of the source terms in the equation~\eqref{eq:cons} is omitted for a more compact expression.
Here, $\bm{q}$ are the conserved quantities:
\begin{align}
	q_{\mathcal{E}}  =&  \psi^6 \sqrt{\bar{\gamma}/\hat{\gamma}} \mathcal{E} \\
	q_{\mathcal{F}_{j}} =&  \psi^6 \sqrt{\bar{\gamma}/\hat{\gamma}} \mathcal{F}_{j}
\end{align}
The corresponding fluxes $\bm{f}^i$ are given by:
\begin{align}
	\left(f_{\mathcal{E}}\right)^i =& \psi^6 \sqrt{\bar{\gamma}/\hat{\gamma}} \left[\alpha {\mathcal{F}}^i - \mathcal{E} \beta^i \right] , \\
	\left(f_{ {\mathcal{F}}_j}\right)^i =& \psi^6 \sqrt{\bar{\gamma}/\hat{\gamma}} \left[  \alpha {\mathcal{P}}^i_{\; j} - {\mathcal{F}}_j \beta^i \right].
\end{align}
The fluxes in the frequency-space are:
\begin{align}
	&\begin{aligned}\label{eq:flux_e}
		{f_\nu}_{\mathcal{E}} = & \alpha \psi^6 \sqrt{\bar{\gamma}/\hat{\gamma}} \left[ n_\mu \mathcal{U}^{\mu\nu\rho} \nabla_\rho u_\nu \right] \\
		 = & \psi^6 \sqrt{\bar{\gamma}/\hat{\gamma}} \nu \Bigg\{ 
		W\Bigg[ \left( \mathcal{Z} v^i - \mathcal{Y}^i \right){\partial_i \alpha}
		- \mathcal{Y}_k v^i {\partial_i \beta^k} \\
		& \qquad - \alpha \mathcal{X}^{ki} \left( \frac{1}{2} v^m \partial_m \gamma_{ki} - K_{ki}\right) \Bigg] \\
		& + \Bigg[ \left[ \mathcal{Z} \partial_t W - \mathcal{Y}_k \partial_t \left(Wv^k\right) \right] 
		 + \left[ \alpha \mathcal{Y}^i - \mathcal{Z} {\beta^i} \right] \partial_i W \\
		& \qquad - \left[ \alpha \mathcal{X}_k^{\;\;i} - \mathcal{Y}_k {\beta^i} \right] \partial_i \left(Wv^k\right) 
		\Bigg] \Bigg\} ,
	\end{aligned} 
	\\
	&\begin{aligned}\label{eq:flux_f}
		{f_\nu}_{\mathcal{F}_{j}} = & \alpha \psi^6 \sqrt{\bar{\gamma}/\hat{\gamma}} \left[ - \gamma_{i \mu} \mathcal{U}^{\mu\nu\rho} \nabla_\rho u_\nu \right] \\
		= & \psi^6 \sqrt{\bar{\gamma}/\hat{\gamma}} \nu \Bigg\{ 
		W\Bigg[ \left( \mathcal{Y}_j v^i - \mathcal{X}_j^{\;\;i} \right){\partial_i \alpha}
		 - \mathcal{X}_{jk} v^i {\partial_i \beta^k} \\
		& \qquad - \alpha \mathcal{W}_j^{\;\;ki} \left( \frac{1}{2} v^m \partial_m \gamma_{ki} - K_{ki}\right) \Bigg] \\
		& + \Bigg[ \left[ \mathcal{Y}_j \partial_t W - \mathcal{X}_{jk} \partial_t \left(Wv^k\right) \right] 
		+ \left[ \alpha \mathcal{X}_j^{\;\;i} - \mathcal{Y}_j {\beta^i} \right] \partial_i W \\
		& \qquad - \left[ \alpha \mathcal{W}_j^{\;\;ki} - \mathcal{X}_{jk} {\beta^i} \right] \partial_i \left(Wv^k\right) 
		\Bigg] \Bigg\} ,
	\end{aligned} 
\end{align}
{where $v^i$ is the fluid 3-velocity and  $W \equiv 1/\sqrt{1-v^i v_i}$ is the Lorentz factor.}
For the details of the derivation, we refer readers to \cite{2013PhRvD..87j3004C, 2020LRCA....6....4M}.

The corresponding gravitational source terms $\bm{s}_{\text{grav}}$ are given by:
\begin{align}
	{s_{\text{grav}}}_{\mathcal{E}} = & \psi^6 \sqrt{\bar{\gamma}/\hat{\gamma}} \Big\{ - {\mathcal{F}}^j \hat{\nabla}_j \alpha  + {\mathcal{P}}^{ij}K_{ij} \Big\} , \\
	{s_{\text{grav}}}_{{\mathcal{F}}_i} = & \psi^6 \sqrt{\bar{\gamma}/\hat{\gamma}} \Big\{
                  - \mathcal{E} \hat{\nabla}_i \alpha + {\mathcal{F}}_k \hat{\nabla}_i \beta^k 
                  + \frac{1}{2} \alpha {\mathcal{P}}^{jk} \hat{\nabla}_i \gamma_{jk} \Big\} , 
\end{align}
where $K_{ij}$ is the extrinsic curvature.
The only non-vanishing geometrical source terms $\bm{s}_{\text{geom}}$ arise for the evolution equation of ${\mathcal{F}}_i$ is 
\begin{align}
	{s_{\text{geom}}}_{\mathcal{E}} = & 0 ,\\ 
	{s_{\text{geom}}}_{{\mathcal{F}}_i} = & \hat{\Gamma}^l_{ik} \left( f_{ {\mathcal{F}}_l} \right)^k,
\end{align}
where the 3-Christoffel symbols $ \hat{\Gamma}^l_{ik} $ associated with the reference metric $\hat{\gamma}_{ij}$.
Finally, the radiation-matter coupling source terms $\bm{s}_{\text{rad}}$ are given by
\begin{align}
	{s_{\text{rad}}}_{\mathcal{E}} = &- \alpha \sqrt{{\gamma}/\hat{\gamma}} \mathcal{S}^\mu_{\text{rad}} n_\mu \label{eq:nu_e_source} ,\\
	{s_{\text{rad}}}_{{\mathcal{F}}_i} = &\alpha \sqrt{{\gamma}/\hat{\gamma}} \mathcal{S}^\mu_{\text{rad}} \gamma_{i\mu}\label{eq:nu_f_source}.
\end{align}

\subsection{Coupling to the hydrodynamical and metric equations}
The radiation fields contribute to the total energy momentum tensor, which affects the hydrodynamical and metric equations.
The hydrodynamical evolution equations are essentially the same as in \cite{2021MNRAS.508.2279C, 2022ApJS..261...22C}, except that the radiation four-force also arise in the source terms of energy and momentum equations, namely:
\begin{align}
        s_{\tau} &\rightarrow s_{\tau} - \sum_{\rm{species}} \int {s_{\text{rad}}}_{\mathcal{E}} \dd{V_\nu},  \\
        s_{S_i} &\rightarrow s_{S_i} - \sum_{\rm{species}} \int {s_{\text{rad}}}_{{\mathcal{F}}_i}  \dd{V_\nu}.
\end{align}

To consistently solve the metric equations, the contribution of the radiations must be taken into account as well.
This can be done simply by including the $3+1$ decomposed source terms for radiation into our metric solver \citep{2020CQGra..37n5015C, 2021MNRAS.508.2279C}.
The $3+1$ decomposed source terms for radiation can be obtained by
\begin{align}
        U_{\rm{rad}} = & \sum_{\rm{species}} \left\{ n_\mu n_\nu T^{\mu\nu}_{\text{rad}} \right\},  \\
        S^i_{\rm{rad}} = & \sum_{\rm{species}} \left\{ - n_\mu \gamma^i_\nu T^{\mu\nu}_{\text{rad}} \right\},  \\
        S_{\rm{rad}} = & \sum_{\rm{species}} \left\{ \gamma^i_\mu \gamma^j_\nu T^{\mu\nu}_{\text{rad}} \right\}, 
\end{align}
where $T^{\mu\nu}_{\text{rad}}$ is the frequency-integrated energy-momentum tensor (see equation~\eqref{eq:em_tensor}) of the corresponding type of radiation.

\section{Numerical methods}\label{sec:numerical_methods} 
\subsection{\label{sec:discretisation}Discretisation}
The discretisation in the spatial and frequency-space for all quantities is based on the finite-volume approach.

An orthogonal system of coordinates $\left(x^1, x^2, x^3 \right)$ is discretised as follows.
The computational domain is divided into $N_1 \times N_2 \times N_3$ cells, where each cell can be represented with a vector of integer numbers $(\texttt{i},\texttt{j},\texttt{k})$ and $1\leq \texttt{i} \leq N_1$, $1\leq \texttt{j} \leq N_2$ and $1\leq \texttt{k} \leq N_3$.
The cell bounds are given by $\left( x^1_\texttt{i-1/2}, x^1_\texttt{i+1/2}\right)$, $\left( x^2_\texttt{j-1/2}, x^2_\texttt{j+1/2}\right)$ and $\left( x^3_\texttt{k-1/2}, x^3_\texttt{k+1/2}\right)$, respectively.
In other words, the mesh spacings can be represented as
\begin{equation}
	\begin{aligned}
		\Delta x^1_\texttt{i} =& x^1_\texttt{i+1/2} - x^1_\texttt{i-1/2}, \\
		\Delta x^2_\texttt{j} =& x^2_\texttt{j+1/2} - x^2_\texttt{j-1/2}, \\
		\Delta x^3_\texttt{k} =& x^3_\texttt{k+1/2} - x^3_\texttt{k-1/2},
	\end{aligned}
\end{equation}
with the cell centre
\begin{equation}
	\begin{aligned}
		x^1_\texttt{i} =& \frac{1}{2} \left( x^1_\texttt{i+1/2} + x^1_\texttt{i-1/2} \right) , \\
		x^2_\texttt{j} =& \frac{1}{2} \left( x^2_\texttt{j+1/2} + x^2_\texttt{j-1/2}\right) , \\
		x^3_\texttt{k} =& \frac{1}{2} \left( x^3_\texttt{k+1/2} + x^3_\texttt{k-1/2}\right) ,
	\end{aligned}
\end{equation}
The cell volume and the surface area, which are associated with the reference metric $\hat{\gamma}_{ij}$, are defined as 
\begin{align}
	&\Delta V \equiv \int_{\text{cell}} \sqrt{\hat{\gamma}} \dd{x^1} \dd{x^2} \dd{x^3} ,\\
	&\Delta A^i \equiv \int_{\text{surface}} \sqrt{\hat{\gamma}} \dd{x^{j,j\neq i}} .
\end{align}
For the calculation of the cell volume $\Delta V$, surface $\Delta A$ and the 3-Christoffel symbols $ \hat{\Gamma}^l_{ik} $, we refer readers to the appendix section in \cite{2021MNRAS.508.2279C}.

Additionally, the frequency-space is discretised by $N_\nu$ frequency bins, where each bin can be represented with an integer $ 1 \leq \texttt{f} \leq N_\nu$ and the corresponding bounds are given by $\left( \nu_\texttt{f-1/2}, \nu_\texttt{f+1/2}\right)$.
The mesh spacing in the frequency-space can be written as
\begin{equation}\label{eq:e_face}
	\Delta \nu_\texttt{f} = \nu_\texttt{f+1/2} - \nu_\texttt{f-1/2}, 
\end{equation}
with the cell centre
\begin{equation}\label{eq:e_centre}
	\nu_\texttt{f} = \frac{1}{2} \left( \nu_\texttt{f+1/2} + \nu_\texttt{f-1/2} \right). 
\end{equation}
In most of the cases, the frequency bins are logarithmically spaced.
Given the upper and lower bounds of the frequency bins $\nu_{\max} \equiv \nu_{N_\nu + 1/2}$ and $\nu_{\min} \equiv \nu_{1/2}$, and the number of frequency bins $N_\nu$, the frequency-space can be discretised as the following.
The $\Delta \nu$ for the first frequency bin (at \texttt{f} = \texttt{1}) can be obtained by
\begin{equation}
	\Delta \nu_\texttt{1} = \left(\nu_{\max}-\nu_{\min}\right) \left(\frac{1-q}{1-q^{N_\nu}}\right),
\end{equation}
where 
\begin{equation}
	q \equiv \left(\frac{\nu_{\max}}{\nu_{\min}}\right)^{1/N_\nu}
\end{equation}
is the scale factor.
The rest of the $\Delta \nu$ can be obtained by the recursion relation
\begin{equation}
	\Delta \nu_{\texttt{f}} = q \Delta \nu_{\texttt{f-1}}.
\end{equation}
With the relation between the width of the frequency bin $\Delta \nu$ and also the corresponding cell interface and centre (see equations~\eqref{eq:e_face} and \eqref{eq:e_centre}), the grid of the frequency-space can be generated.
The cell volume and the surface area in the one-dimensional (spherically symmetric) frequency-space are
\begin{align} \label{eq:rad_dV}
	&\Delta V_\nu \equiv \int_{\text{cell}} 4 \pi \nu^2 \dd{\nu}, \quad 
	\Delta A_\nu \equiv 4 \pi \nu^2 .
\end{align}

In \texttt{Gmunu}, the radiation quantities are volume-averaged in the spatial space and ``frequency-integrated'' in the momentum space.
In particular, the quantity $\left< \bm{q} \right>$ at the centroid $\left({\texttt{i,j,k,f}}\right)$ and the cell interface $\left({\texttt{i+1/2,j,k,f}}\right)$ can be expressed as
\begin{equation}
	\left< \bm{q} \right>_{\texttt{i,j,k,f}} \equiv \frac{1}{\Delta V_{\texttt{i,j,k}}} \int_{\Delta V_{\texttt{i,j,k}}} \dd{V}
		\int_{\Delta {V_\nu}_{\texttt{f}}} \dd{V_\nu} \bm{q},
\end{equation}
and
\begin{equation}
	\left< \bm{q} \right>_{\texttt{i+1/2,j,k,f}} \equiv \frac{1}{\Delta A_{\texttt{i+1/2,j,k}}} \int_{\Delta A_{\texttt{i+1/2,j,k}}} \dd{A}
		\int_{\Delta {V_\nu}_{\texttt{f}}} \dd{V_\nu} \bm{q},
\end{equation}
respectively.

\subsection{Higher moments}
Since only first two moments (the zeroth- and first-moment $\mathcal{E}$ and $\mathcal{F}_i$) are evolved (see equation~\eqref{eq:evolution_e} and \eqref{eq:evolution_f}) while the higher moments such as $\mathcal{P}^{\mu\nu}$ and $\mathcal{U}^{\mu\nu\rho}$ in general cannot be expressed in terms of $\mathcal{E}$ and $\mathcal{F}_i$, a closure relation for determining the higher moments is needed to close the whole system.

In this work, we adopt the (approximate) analytic closure which combines the optically thin and optically thick limits
\begin{equation}
	\mathcal{P}^{\mu\nu} = d_{\text{thin}} \mathcal{P}^{\mu\nu}_{\text{thin}} + d_{\text{thick}} \mathcal{P}^{\mu\nu}_{\text{thick}}, 
\end{equation}
where $\mathcal{P}^{\mu\nu}_{\text{thin}}$ and $\mathcal{P}^{\mu\nu}_{\text{thick}}$ are the Eulerian frame radiation pressure tensors in the optically thin and thick limit respectively.
Here, we have defined
\begin{equation}
	d_{\text{thin}} \equiv \frac{1}{2} \left( 3 \chi - 1\right) ; \;\;
	d_{\text{thick}} \equiv \frac{3}{2} \left( 1 - \chi \right), 
\end{equation}
where $\chi \in \left[ \frac{1}{3}, 1\right]$ is the Eddington factor.
Similarly, the third moment in the fluid frame, which is needed to compute the energy advection term, can be expressed as
\begin{align}
	\mathcal{L}^{\mu\nu\rho} &= d_{\text{thin}} \mathcal{L}^{\mu\nu\rho}_{\text{thin}} + d_{\text{thick}} \mathcal{L}^{\mu\nu\rho}_{\text{thick}}.
\end{align}

In the optically thin limit, the radiation pressure tensor $\mathcal{P}_{\mu\nu}$ in the Eulerian frame is chosen to be (see \cite{2011PThPh.125.1255S})
\begin{equation}\label{eq:Pij}
	\mathcal{P}^{\mu\nu}_{\text{thin}} = \mathcal{E}\frac{\mathcal{F}^\mu \mathcal{F}^\nu}{ \mathcal{F}^i \mathcal{F}_i}.
\end{equation}
while the corresponding fluid frame third moment is
\begin{equation}
	\mathcal{L}^{\mu\nu\rho}_{\text{thin}} = \mathcal{J} \frac{\mathcal{H}^\mu \mathcal{H}^\nu \mathcal{H}^\rho}{\left(\mathcal{H}^2\right)^{3/2}}.
\end{equation}

On the other hand, in the optically thick limit, where the fluid and radiation are in equilibrium, the radiation field is isotropic in the comoving frame
\begin{equation}
	\mathcal{K}^{\mu\nu}_{\text{thick}} = \frac{1}{3} \mathcal{J} h^{\mu\nu} ,
\end{equation}
where $h_{\mu\nu} = g_{\mu\nu} + u_{\mu} u_{\nu}$.
Correspondingly, the radiation pressure tensor $\mathcal{P}_{\mu\nu}$ in the Eulerian frame is
\begin{equation}
	\begin{aligned}
		\mathcal{P}^{\mu\nu}_{\text{thick}} = &
			\frac{4}{3} \mathcal{J} \left( W v^\mu\right) \left( W v^\nu \right) + \frac{1}{3} \mathcal{J} \gamma^{\mu\nu} \\
			& + \left( \gamma^{\mu}_\alpha \mathcal{H}^\alpha \right) \left( W v^\nu \right) 
			+ \left( \gamma^{\nu}_{\beta} \mathcal{H}^\beta \right) \left( W v^\mu \right) ,
	\end{aligned}
\end{equation}
which can be expressed in the terms of the variables in the Eulerian frame by
\begin{align}
	&\frac{\mathcal{J}}{3} = \frac{1}{2W^2+1} \left[ \mathcal{E}\left( 2W^2 - 1 \right) - 2W^2 \mathcal{F}^i v_i \right],\\
	&\begin{aligned}
	\gamma^\alpha_\beta \mathcal{H}^\beta 
		= & \frac{\mathcal{F}^\alpha}{W} + \frac{Wv^\alpha}{2W^2+1} \left[ \left( 4W^2+1\right) \mathcal{F}^i v_i - 4 W^2 \mathcal{E} \right] \\
		= & \frac{\mathcal{F}^\alpha}{W} + Wv^\alpha \left[ \mathcal{F}^i v_i - \mathcal{E} - \frac{\mathcal{J}}{3} \right] .
	\end{aligned}
\end{align}
The fluid frame third moment in the optically thick limit is
\begin{equation}
	\mathcal{L}^{\mu\nu\rho}_{\text{thick}} = \frac{1}{5} \left( \mathcal{H}^\mu h^{\nu\rho} + \mathcal{H}^\nu h^{\rho\mu} + \mathcal{H}^\rho h^{\mu\nu} \right).
\end{equation}

Note that the equation~\eqref{eq:Pij} is derived by assuming the radiation is symmetric around the direction parallel to the flux \citep{2017MNRAS.469.1725M}.
Although this assumption is valid in spherical symmetry, this is not guaranteed in general cases.
For instance, while this relation is asymptotically correct in the optically thick region, this is in general not the case in the free-streaming region because the radiation in vacuum are not all propagating in the same direction \citep{2022arXiv220902538F}.
As a result, this approach fails to describe crossing radiation beams (see e.g. \cite{2013MNRAS.429.3533S, 2015PhRvD..91l4021F, 2020MNRAS.495.2285W, 2022arXiv220902538F}).

\subsection{Closure relation}
A closure relation is needed to compute the Eddington factor $\chi$.
The choice of closure relation affects the accuracy of the two-moment solution.
For more formal discussion and comparison of different analytic closure relations, we refer readers to \cite{2017MNRAS.469.1725M, 2018MNRAS.475.4186F, 2020PhRvD.102h3017R}.
In this work, we adopt the maximum-entropy closure \citep{1978JQSRT..20..541M}, which is given by
\begin{equation}\label{eq:closure}
	\chi \left( \zeta \right) = \frac{1}{3} + {\zeta^2}\frac{2}{15} \left( 3 - \zeta + 3 \zeta^2 \right),
\end{equation}
where the flux factor $\zeta$ is defined as
\begin{equation}\label{eq:flux_factor}
	\zeta \equiv \sqrt{{\mathcal{H}^\mu \mathcal{H}_\mu}/{\mathcal{J}^2}}.
\end{equation}
In the optically thin limit, $\zeta \approx 1$ and thus $\chi \approx 1$.
Conversely, in the optically thick limit, $\zeta \approx 0$ and thus $\chi \approx 1/3$.

Since the flux factor $\zeta$ is defined by the fluid frame moments $\mathcal{J}$ and $\mathcal{H}_\mu$ instead of the observer frame moments $\mathcal{E}$ and $\mathcal{F}_\mu$, the computation of $\zeta$ requires a root-finding process.
As in \cite{2015PhRvD..91l4021F, 2020MNRAS.495.2285W}, we numerically find the root of 
\begin{equation}\label{eq:master_eq}
	f\left( \zeta \right) = \frac{\zeta^2 \mathcal{J}^2 - \mathcal{H}^\mu \mathcal{H}_\mu}{\mathcal{E}^2}.
\end{equation}
Since $f\left( \zeta \right) $ is smooth and its derivative can be expressed analytically, we numerically solve equation~\eqref{eq:master_eq} with the Newton-Raphson method, which is usually more efficient than bracketing methods.
In case the Newton-Raphson method fails to converge, we use the Brent-Dekker method to solve this equation in the range $\zeta \in \left[0,1\right]$.

\subsection{\label{sec:adv_space}Advection in space}
The numerical method for computing the fluxes for space advection is essentially the same as the high-resolution shock-capturing method, except that a slightly modified Harten, Lax and van Leer (HLL) Riemann solver \cite{harten1983upstream} is used.
As pointed out by multiple authors (e.g. \cite{2015ApJS..219...24O, 2015PhRvD..91l4021F, 2016ApJS..222...20K, 2019ApJS..241....7S, 2020MNRAS.495.2285W, 2022MNRAS.512.1499R}), the standard HLL Riemann solver would work only for the optically thin limit, while it fails to reduce the asymptotic diffusion limit when the opacity (absorption plus scattering) is large.
To recover the asymptotic diffusion limit, in this work, we adopt the modification proposed in \cite{2002astro.ph..6281A}, which has been applied in \cite{2013ApJ...762..126O,2016ApJS..222...20K}.
To keep the notation compact, we discuss the approach in the $x$-direction for simplicity.
Specifically, the HLL fluxes for $\mathcal{E}$ and $\mathcal{F}_i$ are modified as
\begin{align}\label{eq:hll}
	&\begin{aligned}
	{f_{\mathcal{E}}}^{\text{HLL}}_{\texttt{i+1/2}} = &\frac{
		\lambda_{+}F_{-} - \lambda_{-}F_{+}
		+ \delta_{\texttt{i+1/2}} \lambda_{+}\lambda_{-}\left(q_{+}-q_{-}\right)
		}{\lambda_{+}-\lambda_{-}}
	\end{aligned}, \\
	&\begin{aligned}
	{f_{ {\mathcal{F}}_j}}^{\text{HLL}}_{\texttt{i+1/2}} = &\frac{
		\delta_{\texttt{i+1/2}}^2 \left(\lambda_{+}F^j_{-} - \lambda_{-}F^j_{+}\right)
		+ \delta_{\texttt{i+1/2}}^2 \lambda_{+}\lambda_{-}\left(q^j_{+}-q^j_{-}\right)
		}{\lambda_{+}-\lambda_{-}} \\
		& + \left( 1 - \delta_{\texttt{i+1/2}}^2 \right)\frac{F^j_{-} + F^j_{+}}{2}
	\end{aligned},
\end{align}
where $\lambda_\pm$ are the characteristic speeds in the $x$-direction.
$\delta_{\texttt{i+1/2}}$ is the newly introduced modification parameter, which is defined as
\begin{equation}
	\delta_{\texttt{i+1/2}} = \tanh \left( \frac{1}{\left(\kappa_{\text{as}}\right)_{\texttt{i+1/2}} \Delta x}\right),
\end{equation}
where $\Delta x$ is the grid width and 
\begin{equation}
	\left(\kappa_{\text{as}}\right)_{\texttt{i+1/2}} = \sqrt{ \left( \kappa_{a} + \kappa_{s}\right)_{\texttt{i}}  \left( \kappa_{a} + \kappa_{s} \right)_{\texttt{i+1}} }
\end{equation}
is the total opacity at cell interface of index $\texttt{i+1/2}$.
In the optically thin region, the modification parameter $\delta \approx 1$ so that the modified flux~\eqref{eq:hll} reduces to the standard HLL flux.
Conversely, the numerical dissipation term vanishes in the optically thick region as the modification parameter $\delta \ll 1$.

The characteristic speeds along the $i$ direction are given by the interpolation of the characteristic speeds between the optically thin and thick limit (see \cite{2011PThPh.125.1255S}) 
\begin{equation}
	\lambda^i_\pm = d_{\text{thin}} \lambda^i_{\pm,\text{thin}} + d_{\text{thick}} \lambda^i_{\pm,\text{thick}},
\end{equation}
where the characteristic speeds in the optically thin and thick limits are given by
\begin{align}
	\lambda^i_{\pm,\text{thin}} = & -\beta^i \pm \alpha \frac{ \left| \mathcal{F}^i \right| }{\sqrt{\mathcal{F}^j\mathcal{F}_j}},\\
	\lambda^i_{\pm,\text{thick}} = & -\beta^i + \frac{2W^2 p^i \pm r }{2W^2 + 1},
\end{align}
with $r \equiv \sqrt{\alpha^2 \gamma^{ii}\left( 2W^2+1 \right) - 2 \left( W p^i \right)^2 }$ and $p^i \equiv \alpha v^i / W $.
Note that, to prevent superluminal characteristic speed, \texttt{Gmunu} reconstructs $\left( \mathcal{E}, \mathcal{F}_i/\mathcal{E}\right)$ instead of $\left( \mathcal{E}, \mathcal{F}_i\right)$ \citep{2018ApJ...854...63O, 2020MNRAS.495.2285W}.

\subsection{\label{sec:energy_advection}Advection in frequency-space}
The computation of the fluxes in frequency-space $\bm{f_\nu}$ (equations~\eqref{eq:flux_e} and \eqref{eq:flux_f}) require the Eulerian decomposed variables of the third-rank moment $\mathcal{U}^{\mu\nu\rho}$, which can be obtained by (see \cite{2013PhRvD..87j3004C, 2020LRCA....6....4M})
\begin{align}
	\mathcal{W}^{\mu\nu\rho} =& \gamma^{\mu}_{\;\;\sigma}\gamma^{\nu}_{\;\;\kappa}\gamma^{\rho}_{\;\;\lambda} \mathcal{U}^{\sigma\kappa\lambda}, \\
	\mathcal{X}^{\mu\nu} =& \frac{\mathcal{S}^{\mu\nu}}{W} + v_\rho \mathcal{W}^{\mu\nu\rho}, \\
	\mathcal{Y}^{\mu} =& \frac{\mathcal{F}^{\mu}}{W} + v_\nu \mathcal{X}^{\mu\nu}, \\
	\mathcal{Z} =& \frac{\mathcal{E}}{W} + v_\mu \mathcal{Y}^{\mu} = \alpha^3 \mathcal{U}^{ttt},
\end{align}
where the third-rank moment $\mathcal{U}^{\mu\nu\rho}$ can be obtained by equation~\eqref{eq:uabc} with the fluid-frame moments $\left\{ \mathcal{J}, \mathcal{H}^{\mu}, \mathcal{K}^{\mu \nu}, \mathcal{L}^{\mu \nu \rho} \right\}$.

Note that fluid accelerations (i.e. time-derivatives of the fluid velocities) involve in the fluxes in frequency-space $\bm{f_\nu}$ (equations~\eqref{eq:flux_e} and \eqref{eq:flux_f}).
The terms that proportional to the fluid accelerations are effectively of the order of $\mathcal{O}\left( v^2/c^2\right)$ in the radiation transport equations in the comoving frame \citep{1979JQSRT..22..293B, 1984Ap&SS.107..333K, 1986CoPhR...3..127M, 2015MNRAS.453.3386J, 2001JQSRT..69..291L, 2002A&A...396..361R}.
{
The radiation transport equations in the comoving frame are correct up to the order of $\mathcal{O}\left( v/c\right)$ if these terms are ignored \citep{2015MNRAS.453.3386J, 2018ApJ...854...63O, 2019ApJS..241....7S}.
In the current implementation, the time-derivatives of the Lorentz factor $\partial_t \left(W\right)$ and the velocities $\partial_t \left(Wv^i\right)$ are calculated simply by first-order backward differencing with the values of the previous time step, similar to \cite{2015ApJS..219...24O}.
}
Adding these terms while preserving numerical stabilities is non-trivial, the {proper treatment} of these terms will be investigated in a future study.

Similar to the advection in space described in section~\ref{sec:adv_space}, the frequency advection term integrated with a frequency cell $\dd{V_\nu}$ at the $\texttt{f}$-th frequency cell can be calculated by
\begin{equation}\label{eq:numerical_advection}
	\begin{aligned}
		& \left[\int_{\Delta{V_\nu}} \frac{1}{\nu^2}\partial_\nu \left[ \nu^2 \bm{f_\nu} \right] \dd{V_\nu} \right]_{\texttt{f}} \\
		& = 
		\left[ \left( \left<\bm{f_\nu}\right>\Delta A_\nu\right)\Big|_{\texttt{f+1/2}} -  \left( \left<\bm{f_\nu}\right>\Delta A_\nu \right) \Big|_{\texttt{f-1/2}} \right] ,
	\end{aligned}
\end{equation}
where the cell surface area is given by equation~\eqref{eq:rad_dV}.
As stated in section~\ref{sec:discretisation}, the advection term is integrated with the corresponding frequency bin since \texttt{Gmunu} manipulates frequency-bin-integrated radiation quantities.
Currently, \texttt{Gmunu} handles this energy advection term explicitly.

Since the fluxes in frequency-space $\bm{f_\nu}$ can be expressed in terms of linear combinations of the fluid-frame radiation momenta $\left\{ \mathcal{J}, \mathcal{H}^{\mu}, \mathcal{K}^{\mu \nu}, \mathcal{L}^{\mu \nu \rho} \right\}$, the energy and momentum are conserved as long as the fluxes vanish at the outer boundary in the frequency space.
Similar to \cite{2010ApJS..189..104M, 2016ApJS..222...20K}, we split the flux as
\begin{equation}
	\left<\bm{f_\nu}\right>_{\texttt{f+1/2}} \equiv \left<\bm{f_\nu}\right>^{\text{L}}_{\texttt{f}} + \left<\bm{f_\nu}\right>^{\text{R}}_{\texttt{f+1}},
\end{equation}
where we have defined
\begin{align}
	\left<\bm{f_\nu}\right>^{\text{L}}_{\texttt{f}} \equiv &\left<\bm{f_\nu}\right>_{\texttt{f}} w_{\texttt{f}}, \\
	\left<\bm{f_\nu}\right>^{\text{R}}_{\texttt{f}} \equiv &\left<\bm{f_\nu}\right>_{\texttt{f}} \left( 1 - w_{\texttt{f}} \right),
\end{align}
with the weighting function $w$
\begin{equation}
	w_\texttt{f} \equiv \frac{j^\sigma_\texttt{f+1/2}}{j^\sigma_\texttt{f-1/2} + j^\sigma_\texttt{f+1/2}}.
\end{equation}
Here, $j^\sigma_\texttt{f+1/2}$ is the weighted geometric mean of the distribution function $j$ at cell interface $\texttt{f+1/2}$, which is given by
\begin{equation}
	j^\sigma_\texttt{f+1/2} \equiv \left[
	\left(\frac{\mathcal{J}_\texttt{f}}{\nu_\texttt{f}}\right)^{1-r_\texttt{f+1/2}} 
	\left(\frac{\mathcal{J}_\texttt{f+1}}{\nu_\texttt{f+1}}\right)^{r_\texttt{f+1/2}} \right]^\sigma,
\end{equation}
with $r_\texttt{f+1/2} \equiv { \left( \nu_\texttt{f+1/2} - \bar{\nu}_\texttt{f} \right)}/{\left(\bar{\nu}_\texttt{f+1} - \bar{\nu}_\texttt{f} \right)}$,
where $\bar{\nu}_\texttt{f}$ denotes the centroid of the $\texttt{f}$-th cell.
By default, we use ``Harmonic'' interpolation by setting $\sigma = 1$.

Note that, as discussed in \cite{2020LRCA....6....4M}, this frequency-space advection approach has been developed in the context of \emph{Lagrangian} two-moment schemes \citep{2010ApJS..189..104M} to ensure neutrino number conservation.
However, despite the fact that the fluxes in frequency-space $\bm{f_\nu}$ can be expressed in terms of linear combinations of the fluid-frame radiation momenta $\left\{ \mathcal{J}, \mathcal{H}^{\mu}, \mathcal{K}^{\mu \nu}, \mathcal{L}^{\mu \nu \rho} \right\}$ and the success in the frequency advection and application tests (see section~\ref{sec:energy_adv_tests} and \ref{sec:ccsn_test} below, also see \cite{2015ApJS..219...24O, 2016ApJS..222...20K}), it is still unclear whether the neutrino number conservation is still preserved (up to machine precision) if the same approach is applied directly to the Eulerian two-moment scheme as in \cite{2015ApJS..219...24O, 2016ApJS..222...20K}.
Further investigations and comparisons of different frequency/energy advection schemes are needed, which will be left as future work.

\subsection{\label{sec:source_terms}Radiation-fluid interactions}
The radiation-matter coupling source terms~\eqref{eq:source_terms} can be very large when the opacities are large.
From the numerical point of view, these interaction source terms can become very stiff in the optically thick regime, applying explicit time integration would be inefficient due to the extremely strict constraints on the time steps.
Implicit-explicit Runge-Kutta schemes (e.g. \cite{ASCHER1997151, pareschi2005implicit}) offer an effective approach to overcome this challenge.
These schemes have been applied and tested previously in \texttt{Gmunu} for resistive magnetohydrodynamics \citep{2022ApJS..261...22C}, and also in several other radiation hydrodynamics codes (e.g. \cite{2015ApJS..219...24O, 2015PhRvD..91l4021F, 2016ApJS..222...20K, 2020MNRAS.495.2285W, 2020ApJ...900...71A, 2022MNRAS.512.1499R, 2022arXiv221100027I}.
For the details of the implementation of IMEX in \texttt{Gmunu}, we refer readers to our previous work \cite{2022ApJS..261...22C}.

In general, most of the fluid conserved variables $\bm{q}_{\rm{hydro}}$ have to be solved implicitly all together with the radiation moments (see e.g. \cite{2016ApJS..222...20K}).
However, the computational cost is high because one will need to update the primitive variables of fluid (such as pressure and specific energy) during the iteration when tabulated equations of state are being used.
In this work, as in \cite{2015ApJS..219...24O, 2015PhRvD..91l4021F, 2022MNRAS.512.1499R}, we implicitly solve radiation moments $\left\{ q_{\mathcal{E}},q_{\mathcal{F}_{j}} \right\}$ only, and the coupling to the fluid is treated explicitly (see section~\ref{sec:coupling}).
Investigations of more advance fully implicit treatments such as \cite{2019ApJS..241....7S, 2021ApJS..253...52L} are left as future work.

An implicit step which updates the solution of the radiation moments $\left\{ q_{\mathcal{E}},q_{\mathcal{F}_{j}} \right\}$ from the time-step $\texttt{n}$ (which is denoted as $\bm{q}^{\texttt{n}}$) to the next time-step $\texttt{n+1}$ (which is denoted as $\bm{q}^{\texttt{n+1}}$) can be expressed as
\begin{equation}
	\bm{q}^{\texttt{n+1}} = \bm{q}^{\texttt{n}} + \Delta t \bm{s}_{\text{rad}}\left(\bm{q}^{\texttt{n+1}}\right).
\end{equation}
To obtain the updated solution $\bm{q}^{\texttt{n+1}}$, we solve the non-linear system $\bm{f} \left( \bm{q} \right) $, which is defined as
\begin{equation}
	\bm{f} \left( \bm{q} \right) \equiv - \bm{q} + \bm{q}^{\texttt{n}} + \Delta t \bm{s}_{\text{rad}}\left(\bm{q}\right).
\end{equation}
Currently, we solve this non-linear system by using multidimensional Broyden method.
The Jacobian ${\partial f_i}/{\partial q_j}$ of $f_i(q_j)$ is obtained numerically by forward differencing.
The implementation of Broyden solver and computation of Jacobian follows \cite{press1996numerical}.

\subsubsection{\label{sec:initial_guess}Initial guess}
A proper initial guess is needed for the implicit step.
In this work, we follow the approach introduced by \cite{2022MNRAS.512.1499R}.
Although only monochromatic source terms (i.e. equation~\eqref{eq:e_and_a_source} and \eqref{eq:iso_scattering_source}) are considered in this approach, we find that this method usually provides good initial guess.
For completeness, we describe the procedure of obtain the initial guess at a given radiation frequency $\nu$.
First, we transform the solution $\bm{q}^{\texttt{n}}$ into fluid frame, and denote it as $\left\{\tilde{\mathcal{J}}, \tilde{\mathcal{H}}_i \right\}$, and then update the fluid-frame moments by (see the \emph{Lagrangian} two-moment model in \cite{2020LRCA....6....4M})
\begin{align}
	\hat{\mathcal{J}} = &\tilde{\mathcal{J}} + \Delta t \frac{\alpha}{W} \left( \eta - \kappa_a \hat{\mathcal{J}} \right), \\
	\hat{\mathcal{H}}_i = &\tilde{\mathcal{H}}_i - \Delta t \frac{\alpha }{W} \left( \kappa_{a} + \kappa_{s} \right) \hat{\mathcal{H}}_i,
\end{align}
where $\hat{\mathcal{J}}$ and $\hat{\mathcal{H}}_i$ denote the updated fluid-frame moments.
Second, we transform the updated $\hat{\mathcal{J}}$ and $\hat{\mathcal{H}}_i$ into Eulerian frame $\hat{\mathcal{E}}$ and $\hat{\mathcal{F}}_i$ by assuming optically thick
\begin{align}
	\hat{\mathcal{E}} = &\frac{\hat{\mathcal{J}}}{3}\left( 4 W^2 - 1\right) - 2 W \hat{\mathcal{H}}_\alpha n^\alpha, \\
	\hat{\mathcal{F}}_i = & W \hat{\mathcal{H}}_i + \left( \frac{4}{3}W^2\hat{\mathcal{J}} - W \hat{\mathcal{H}}_\alpha n^\alpha \right) v_i, 
\end{align}
where $\hat{\mathcal{H}}_0$ can be computed by the fact that $\hat{\mathcal{H}}_\alpha u^\alpha=0$, and thus $\hat{\mathcal{H}}_\alpha n^\alpha = - \hat{\mathcal{H}}_i v^i$.
The resulting Eulerian-frame moments $\hat{\mathcal{E}}$ and $\hat{\mathcal{F}}_i$ are used as the initial guess of the implicit step.
Here, we assume the optically thick limit since the initial guess becomes important only in the optically thick regime.

Note that, as discussed in \cite{2022MNRAS.512.1499R}, the updated $\hat{\mathcal{J}}$ and $\hat{\mathcal{H}}_i$ are exact solutions only at leading order in $v/c$ where $u^\mu \partial_\mu \approx W \partial_t$ and when only monochromatic source terms are considered.
However, the corresponding Eulerian-frame moments $\hat{\mathcal{E}}$ and $\hat{\mathcal{F}}_i$ are not correct solutions even if the closure is taken into account during the transformation.

\subsubsection{\label{sec:coupling}Coupling to fluid}
Once the radiation moments are solved, we explicitly update the fluid's energy and momentum by
\begin{align}
        q_{\tau} & \rightarrow q_{\tau} - \Delta t \sum_{\rm{species}} \int {s_{\text{rad}}}_{\mathcal{E}} \dd{V_\nu},  \\
        q_{S_i} & \rightarrow q_{S_i} - \Delta t \sum_{\rm{species}} \int {s_{\text{rad}}}_{{\mathcal{F}}_i}  \dd{V_\nu},
\end{align}
where ${s_{\text{rad}}}_{\mathcal{E}}$ and ${s_{\text{rad}}}_{{\mathcal{F}}_i}$ are obtained by equation~\eqref{eq:nu_e_source} and \eqref{eq:nu_f_source} with the updated radiation moments.

\subsubsection{\label{sec:rank}Rank of non-linear system}
In general, a non-linear system of dimensions $(N_{\text{dim}}+1) \times N_\nu \times N_\text{species}$ must be solved.
Here we assume the fluid quantities are kept fixed during the implicit step and consider $N_\nu$ frequency-bins, $N_\text{species}$ species of neutrino, in $N_{\text{dim}}$ dimensional spatial space.
Since the size of the non-linear system could be very large, it is computationally expensive if we fully solve this system.
In practice, depending on the nature of the problem, it is not necessary to apply the full implicit solver.
Avoiding full implicit treatment would significantly reduce the computational cost.
Similar to \cite{2015MNRAS.453.3386J}, we list different modes of the radiation-interaction source terms treatment which are implemented in \texttt{Gmunu}:
\begin{enumerate}[(i)]
	\item \emph{multi-species multi-group}:
		All radiation moments $\left\{ q_{\mathcal{E}},q_{\mathcal{F}_{j}} \right\}$ are solved fully implicitly.
		This is the general mode discussed above, where the dimensions of the non-linear system is $(N_{\text{dim}}+1) \times N_\nu \times N_\text{species}$.
	\item \emph{single-species multi-group}:
		The radiation moments $\left\{ q_{\mathcal{E}},q_{\mathcal{F}_{j}} \right\}$ are solved for each species separately.
		The dimensions of the non-linear system now reduced to $N_\text{species}$ non-linear systems of dimensions $(N_{\text{dim}}+1) \times N_\nu$.
		Since the source terms which contain species coupling are treated explicitly in this mode.
		This mode is less accurate when the species coupling is strong.
	\item \emph{single-species single-group}:
		The radiation moments $\left\{ q_{\mathcal{E}},q_{\mathcal{F}_{j}} \right\}$ are solved for each species and for each frequency-groups separately.
		The dimensions of the non-linear system now reduced to $N_\text{species} \times N_\nu$ non-linear systems of dimensions $(N_{\text{dim}}+1)$.
		Since this mode is purely monochromatic, the coupling of different frequency-group cannot be done implicitly.
		In this mode, only the emission/absorption and elastic scattering source terms ($\mathcal{S}^\mu_{\text{E/A}}$ and $\mathcal{S}^\mu_{\text{ES}}$, see equations~\eqref{eq:e_and_a_source} and \eqref{eq:iso_scattering_source}) are solved implicitly, while the source terms that contain species or frequency couplings are treated explicitly.
		In this case, we have included the analytic Jacobian for implicit solver by following \cite{2022MNRAS.512.1499R}.
		The details of which can be found in appendix~\ref{sec:jaco}.
\end{enumerate}
In practice, to minimise the computational cost, we switch to different mode in different stage of the simulations, where the criteria of which are highly problem dependent.

\subsection{\label{sec:E_to_J}Transformation from Eulerian-frame to fluid-frame}
As shown in the previous sections, although the radiation fields are solved in the Eulerian-frame, the radiation moments in the fluid-frame are often needed in most of the calculations.
The most straight forward way to compute the fluid-frame moments $\left\{ \mathcal{J}, \mathcal{H}^{\mu} \right\} $ is to contract the energy momentum tensor $\mathcal{T}^{\mu\nu}$ with the comoving four-velocities $u^\mu$.
Alternatively, we found it is useful to directly express the fluid-frame moments $\left\{ \mathcal{J}, \mathcal{H}^{\mu} \right\} $ in terms of the Eulerian-frame moments $\left\{ \mathcal{E}, \mathcal{F}^{\mu} \right\} $, especially when only part of the fluid-frame moments are needed.
We decompose $\left\{ \mathcal{J}, \mathcal{H}^{\mu}, \mathcal{H}^{\mu}\mathcal{H}_{\mu} \right\} $ and express them in terms of the Eulerian-frame moments $\left\{ \mathcal{E}, \mathcal{F}^{\mu} \right\} $ by following \cite{spectrecode, 2022MNRAS.512.1499R}.
The details of which can be found in appendix~\ref{sec:decomposed_variables}.

\subsection{Enforcing validity}
Unphysical solutions occasionally arise during the evolution due to the numerical round-off errors especially when the radiation energy density $\mathcal{E}$ is very small.
In \texttt{Gmunu}, we include the following error handling policies to enforce the validity of the numerical solution.

Similar to the standard ``atmosphere'' treatment for rest mass density $\rho$ in hydrodynamical simulation (e.g. \cite{2020CQGra..37n5015C, 2021MNRAS.508.2279C}), we enforce the non-negativity of the energy density $\mathcal{E}$.
In particular, we define a minimum allowed distribution function ${f}_{\min}$ and a threshold ${f}_{\text{thr}}$, where ${f}_{\text{thr}} \geq {f}_{\min} \geq 0$.
Whenever the energy density drop below the threshold (i.e. when $\mathcal{E}\left(\nu\right) < \nu {f}_{\text{thr}}$), we set the energy to be the minimum allowed energy density $\nu {f}_{\min}$, and enforce a vanishing flux by setting $\mathcal{F}_i\left(\nu\right) = 0$.
In the grey transport cases, ${f}_{\min}$ and ${f}_{\text{thr}}$ represent the minimum allowed energy density and the threshold directly.

In addition to the negative energy density, unphysical solutions could also arise when $\mathcal{F}^i\mathcal{F}_i > \mathcal{E}^2$.
Similar to \cite{2016ApJS..222...20K, 2019ApJS..241...28R}, we enforce
\begin{equation}
	\mathcal{F}_i \rightarrow \mathcal{F}_i \times \min\left( \xi_{\max}, \xi_{\max} / \xi\right),
\end{equation}
where we have defined the Eulerian flux factor
\begin{equation}
	\xi \equiv \sqrt{{\mathcal{F}^i \mathcal{F}_i}/{\mathcal{E}^2}},
\end{equation}
and $\xi_{\max}$ is the maximum allowed Eulerian flux factor.
Unless explicitly stated, we set $f_{\rm{thr}} = 10^{-30}$, $f_{\min} = 0$ and $\xi_{\max} = 1$.

\section{Numerical tests}\label{sec:numerical_tests}
In this section, we present a selection of representative test problems with our code to assess the performance and accuracy of our new two-moment-based module.
The tests range from special relativistic to general relativistic radiation transfer, from one to multiple dimensions, and from frequency integrated (grey) to multifrequency group.
Here, we consider only the monochromatic source terms (i.e. at a given radiation frequency $\nu$, the calculation of the radiation emissivity, absorption and scattering coefficients $\eta \left( \nu \right)$, $\kappa_a \left( \nu \right)$ and $\kappa_s \left( \nu \right)$ do not depend on other radiation frequencies $\nu' \neq \nu $, see equation~\eqref{eq:e_and_a_source} and \eqref{eq:iso_scattering_source}) with idealised opacities.
Tests with sophisticated realistic neutrino opacities are presented in section~\ref{sec:application_tests}.

For the frequency-integrated (grey) test, we denote the frequency-integrated radiation energy and momentum in the fluid-frame as
\begin{align}
	J =  \int_0^\infty \mathcal{J} \dd{V_\nu}; \;\;
	H_i = \int_0^\infty \mathcal{H}_i \dd{V_\nu},
\end{align}
and so as the case for the fluid-frame moments
\begin{align}
	E =  \int_0^\infty \mathcal{E} \dd{V_\nu}; \;\;
	F_i =  \int_0^\infty \mathcal{F}_i \dd{V_\nu}.
\end{align}
In addition, we denote the frequency-integrated emissivity $\eta$, energy-averaged absorption and scattering coefficients as
\begin{align}
	\bar{\eta} = & \int_0^\infty \eta \dd{V_\nu}, \\
	\bar{\kappa}_a = & \frac{\int_0^\infty \kappa_a \nu f \dd{V_\nu}}{\int_0^\infty \nu f \dd{V_\nu}}, \\
	\bar{\kappa}_s = & \frac{\int_0^\infty \kappa_s \nu f \dd{V_\nu}}{\int_0^\infty \nu f \dd{V_\nu}},
\end{align}
where $f$ is the distribution function.
Unless otherwise specified, all simulations reported in this paper were performed with Harten, Lax and van Leer (HLL) Riemann solver \cite{harten1983upstream}, 2-nd order Minmod limiter \cite{1986AnRFM..18..337R} with IMEX-SSP2(2,2,2) time integrator \cite{pareschi2005implicit}.

\subsection{Transparent fluid with a velocity jump}
In this section, we consider the propagation of radiation in a moving optically thin medium as in \cite{2022MNRAS.512.1499R}.
In particular, we consider a one-dimensional mildly relativistic fluid moving with Lorentz factor $W=2$ in an opposite direction in a flat spacetime.
The background fluid velocity profile is chosen to be
\begin{align}
	Wv^x = 
	\begin{cases}
		\sqrt{W^2-1}	,&	\text{if } x > 0 \\
		-\sqrt{W^2-1}	,&	\text{otherwise } 
	\end{cases}.
\end{align}
The initial profile of the radiation is set to be
\begin{align}
	{E} = 
	\begin{cases}
		1	,&	\text{if } x < - 0.5 \\
		0	,&	\text{otherwise }
	\end{cases},
\end{align}
and ${F}_x = {E}$.
To consider the case in optically thin limit, we consider zero opacities, i.e. $\bar{\eta} = \bar{\kappa}_s = \bar{\kappa}_a = 0$.
In this test, the hydrodynamical profiles are kept fixed, and disable the interaction between the fluid and radiation during the evolution.
We assume slab geometry, and the computational domain covers the region $\left[ -1, 1 \right]$ with 200 grid points. 

\begin{figure}
	\centering
	\includegraphics[width=\columnwidth, angle=0]{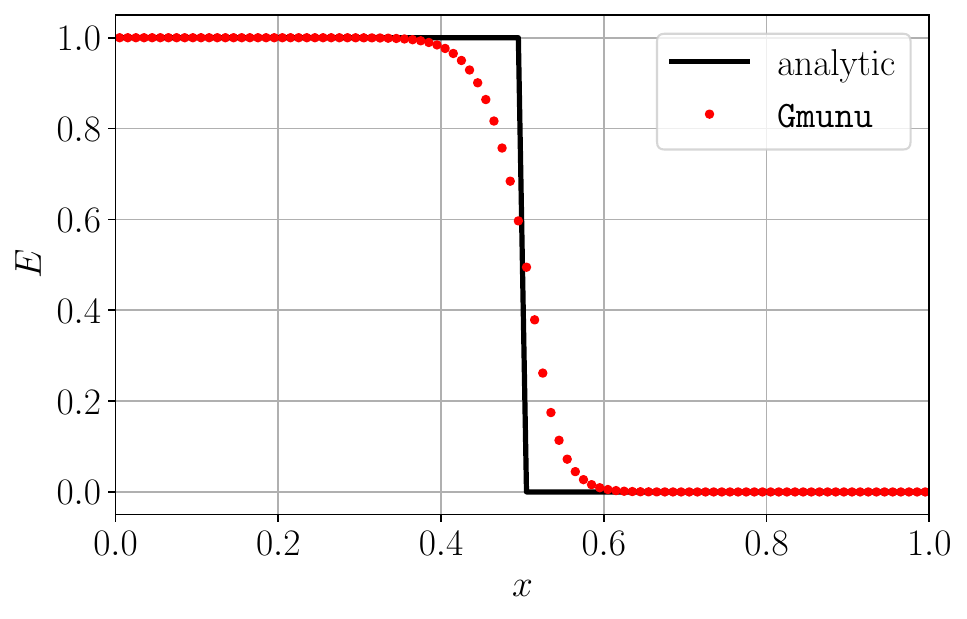}
	\caption{
	The energy density profile of the radiation ${E}$ at $t=1$, where the radiation has propagated through the velocity jump at $x=0$.
	Despite the discontinuity of the velocity profile, the numerical solution obtained by $\texttt{Gmunu}$ does not have artificial oscillations at the velocity jump interface $x=0$ or at the radiation front $x=0.5$.
		}
	\label{fig:M1_velocity_jump_1d}	
\end{figure}

Figure~\ref{fig:M1_velocity_jump_1d} shows the energy density profile of the radiation ${E}$ at $t=1$, where the radiation has propagated through the velocity jump at $x=0$.
Despite the discontinuity of the velocity profile, the numerical solution obtained by \texttt{Gmunu} has artificial oscillations neither at the velocity jump interface $x=0$ nor at the radiation front $x=0.5$.
This test demonstrates that \texttt{Gmunu} is able to handle the radiation transport in a mildly relativistic moving fluid.

\subsection{Homogeneous radiating sphere}
The homogeneous radiating sphere test is a toy model of a hot neutron star which emits neutrinos.
As discussed in \cite{1997A&A...325..203S, 2015ApJS..219...24O, 2017MNRAS.469.1725M, 2020ApJ...900...71A}, this test is sensitive to the chosen closure relation.
In this test, we consider a homogeneous sphere with a radius $R$ with constant emission and absorption rate $\bar{\eta} = \bar{\kappa}_a$ and simply ignore scatterings ($\bar{\kappa}_s = 0$).
As discussed in \cite{1997A&A...325..203S}, this problem has an analytic solution
\begin{equation}
	I(r,\mu) = B \left\{ 1 - \exp\left[ -\bar{\kappa} s\left(r,\mu\right)\right] \right\},
\end{equation}
where $B$ is the strength of the initial energy density of the radiation, 
\begin{align}
	s \left( r,\mu \right) = 
	\begin{cases}
		r \mu + R g \left( r, \mu \right)	&	\text{if } r < R \text{ and } -1 \leq \mu \leq 1, \\
		2 R g\left( r, \mu\right)	&	\text{if } r \geq R \text{ and } \sqrt{1- \left(\frac{R}{r}\right)^2} \leq \mu \leq 1, \\
		0	&	\text{otherwise } ,
	\end{cases}
\end{align}
and
\begin{equation}
	g\left(r, \mu\right) = \sqrt{1 - \left(\frac{r}{R}\right)^2 \left(1-\mu^2\right)},
\end{equation}
in which $\mu \equiv \cos \theta$ is the directional cosine.
The analytic solution for ${J}$ and ${H}$ are
\begin{align}
	\left\{ {J}, {H} \right\} = \frac{1}{2} \int^{1}_{-1} \dd{\mu} \mu^{\left\{0,1\right\}} I.
\end{align}
Note that when the background velocities are zero, the radiation moments in the fluid frame are the same as in Eulerian frame.

We simulate this problem by setting the radius of the sphere $R$ and the strength of the initial energy density of the radiation $B$ to be unity (i.e. $R=1=B$).
The initial profile of the radiation is set to be
\begin{align}
	\left( {E}, {F}_r/{E} \right) = 
	\begin{cases}
		\left( B, 0 \right)	&	\text{if } r < R ,\\
		\left( B \left(\frac{R^2}{r^2}\right), 0.1 \right)	&	\text{otherwise } .
	\end{cases}
\end{align}
The hydrodynamical profiles are kept fixed, and disable the interaction between the fluid and radiation during the evolution.
To see how our code behaves with different opacities in this test, we perform the test with three different absorption opacities: $\bar{\kappa}_a = 10^6, 10, 1$, respectively.
Although this is a spherically symmetric test problem which can be run in one-dimensional spherical coordinates (e.g. \cite{2015ApJS..219...24O}), we simulated this problem in three-dimensional Cartesian coordinates $(x,y,z)$ to minimise the symmetry imposed in the simulation.
In particular, the computational domain covers the region $\left[-5,5\right]\times\left[-5,5\right]\times\left[-5,5\right]$ with the resolution $256^3$.

\begin{figure}
	\centering
	\includegraphics[width=\columnwidth, angle=0]{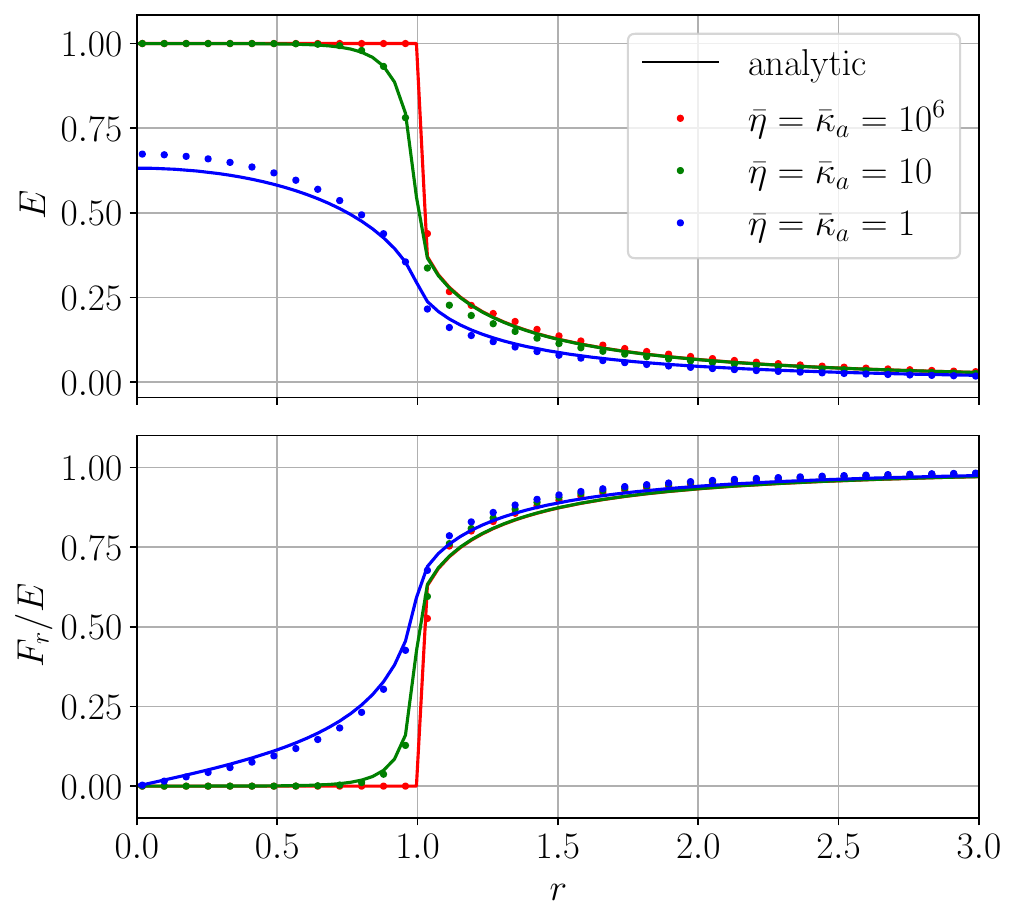}
	\caption{
		Energy density ${E}$ (\emph{upper panel}) and the radial flux ratios ${F}_r/{E}$  (\emph{lower panel}) along $x$-axis at time $t=10$ of the homogeneous radiation sphere test.
		Since the profiles are extracted along the $x$-axis, the $x$ coordinate has the same value of radius $r$.
		The dots show the numerical results obtained by \texttt{Gmunu} while the solid lines show the corresponding analytic solution.
		For the high opacity cases (i.e. $\bar{\kappa}_a \geq 10$, the red and green dots), the numerical results are mostly indistinguishable from the reference solution except the region nearby the discontinuity at the surface of the sphere.
		However, for the low opacity case ($\bar{\kappa}_a = 1$, the blue dots), the numerical results of the energy density $E$ are less accurate inside the sphere.
		This is because the analytic closure does not give the correct second moment in the intermediate opacities regime (i.e. $\kappa \lesssim 1$).
		}
	\label{fig:M1_radiating_sphere_cart_3d}	
\end{figure}

Figure~\ref{fig:M1_radiating_sphere_cart_3d} compares the energy density ${E}$ and the radial flux ratios ${F}_r/{E}$ along $x$-axis at $t=10$ with the analytic solutions.
For the high-opacity cases (i.e. $\bar{\kappa}_a \geq 10$, the red and green dots), \texttt{Gmunu} resolves the optically thick and thin region well despite the discontinuities at the surface of the sphere.
While the analytic closure gives the correct second moment in the high-opacity regime and the free-streaming regime, this is not the case for the intermediate regime (e.g. for the region where the opacity $\kappa \lesssim 1$) \citep{2017MNRAS.469.1725M, 2020MNRAS.495.2285W}.
As a result, for the low opacity ($\bar{\kappa}_a = 1$, blue dots) case, the numerical results of the energy density $E$ inside the sphere is less accurate while the exterior energy density tail still matches the analytic results.
Similar feature has also been seen in \cite{2020MNRAS.495.2285W}.

\subsection{Diffusive limit in a scattering medium}
In this test, we consider the diffusion of radiation when scattering opacity is high and the mean free path is small compare to the grid size $\Delta x$ (e.g. \cite{2022MNRAS.512.1499R,2022arXiv221100027I}).
The initial profile of the radiation is set to be
\begin{align}
	{E} = \theta\left(x+\frac{1}{2}\right) - \theta\left(x-\frac{1}{2}\right),
\end{align}
where $\theta\left(x\right)$ is the Heaviside step function, and ${F}^i = 0$.
We consider this diffusion in a purely scattering medium, we set $\bar{\eta} = 0 = \bar{\kappa}_a$ and $\bar{\kappa}_s = 10^3$.
Here, we again consider static background hydrodynamical profiles, and assume slab geometry.
The computational domain covers the region $\left[ -2, 2 \right]$ with 256 grid points. 

The evolution of the energy density $E$ can be approximated by the diffusion equation $\partial_t E = \left(\partial_x^2 E\right) / {3\kappa_s} $ when the timescales are longer than the equilibrium time \citep{2022MNRAS.512.1499R,2022arXiv221100027I}.
The exact solution of which is given by
\begin{align}\label{eq:diff_sol}
	E\left(t,x\right) = \frac{1}{2}\left[ \erf\left(\frac{x+\frac{1}{2}}{\sqrt{4 \tau t}}\right) - \erf\left(\frac{x-\frac{1}{2}}{\sqrt{4 \tau t}}\right) \right],
\end{align}
where $\tau = 1 / \left(3 \kappa_s \right)$ is the diffusion timescale.

\begin{figure}
	\centering
	\includegraphics[width=\columnwidth, angle=0]{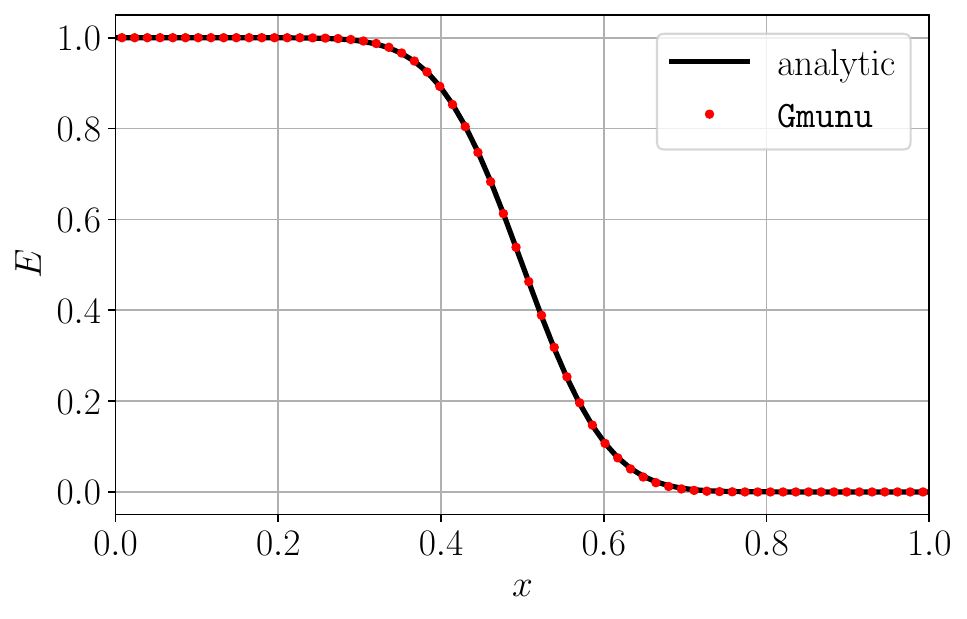}
	\caption{
		The energy density profile of the radiation ${E}$ at $t=10$ in a purely scattering medium with high scattering opacity $\bar{\kappa}_s = 10^3$.
		The red dots show the numerical solution obtained by \texttt{Gmunu} while the black solid line shows the reference analytic solution~\eqref{eq:diff_sol} in the diffusive limit.
		This indicates that \texttt{Gmunu} captures correct diffusion rate even in diffusive limit.
		}
	\label{fig:M1_diffusive_limit_1d}
\end{figure}

Figure~\ref{fig:M1_diffusive_limit_1d} shows the energy density profile of the radiation $E$ at time $t=10$.
As shown in the figure, the numerical solution obtained by \texttt{Gmunu} (red dots) agrees with the reference analytic solution~\eqref{eq:diff_sol} in the diffusive limit (black solid line).
This demonstrates that \texttt{Gmunu} is able to capture correct diffusion rate even when the scattering opacity is high.

\begin{figure}
	\centering
	\includegraphics[width=\columnwidth, angle=0]{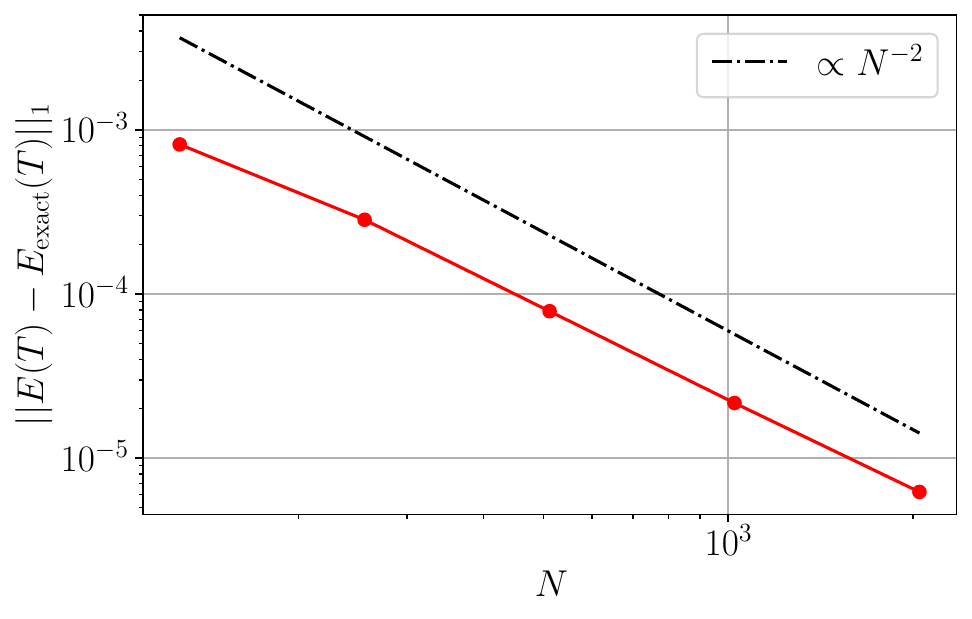}
	\caption{
		The $L_1$-norm of the difference between the exact and final ($t=10$) values of the energy density of the radiation $E$ at different resolution $N$.
		The second-order ideal scaling is given by the black dashed line.
		In this test, second order accurate strong-stability preserving IMEX-SSP2(2,2,2) time integrator \cite{pareschi2005implicit}, Harten, Lax and van Leer (HLL) Riemann solver \cite{harten1983upstream} with 2-nd order Minmod limiter \cite{1986AnRFM..18..337R} are used. 
		The code achieves roughly 1.86 order of convergence in this test.
		}
	\label{fig:M1_diffusive_limit_1d_convergence_rate}
\end{figure}

To quantify the convergence rate at $t=T \equiv 10$, we perform the simulation with different resolutions, and compute the $L_1$-norm of the difference between the exact and final ($t=10$) values of the energy density of the radiation $E$ as
\begin{equation}
	\begin{aligned}
	\lvert\lvert E\left(T\right) - E_{\rm{exact}}\left(T\right)\rvert\rvert_1 \equiv 
		\frac{\sum\limits_i \lvert E(T) - E_{\rm{exact}}(T) \rvert \Delta V_i}{\sum\limits_i \Delta V_i},
	\end{aligned}
\end{equation}
Figure~\ref{fig:M1_diffusive_limit_1d_convergence_rate} shows the $L_1$-norm of the difference between the exact and final ($t=10$) values of the energy density of the radiation $E$.
The order of convergence of this test is roughly 1.86.

\subsection{Diffusive limit in a moving medium}
In this test, we consider a propagation of a radiation in a moving purely scattering medium as in \cite{2022MNRAS.512.1499R,2022arXiv221100027I}.
This is known to be a demanding test that the result is highly sensitive to the implicit treatment \cite{2022MNRAS.512.1499R}.
Consider a Gaussian pulse of radiation
\begin{align}
	{E} = \exp\left( -9 x^2 \right)
\end{align}
which is fully trapped ($H^\mu=0$) in the medium.
The radiation flux in Euler frame can be written as
\begin{align}
	F_i = \frac{4}{3}J W^2 v_i,
\end{align}
where $J = {3E}/\left({4W^2 - 1}\right)$ in this case.
The medium is set to be purely scattering (i.e. $\bar{\eta} = 0 = \bar{\kappa}_a$) with high scattering opacity $\bar{\kappa}_s = 10^3$, which moves with a relativistic velocity $v^x = 0.5$.
Here, we assume slab geometry, and the computational domain covers the region $\left[ -5, 5 \right]$ with 1024 grid points. 

\begin{figure}
	\centering
	\includegraphics[width=\columnwidth, angle=0]{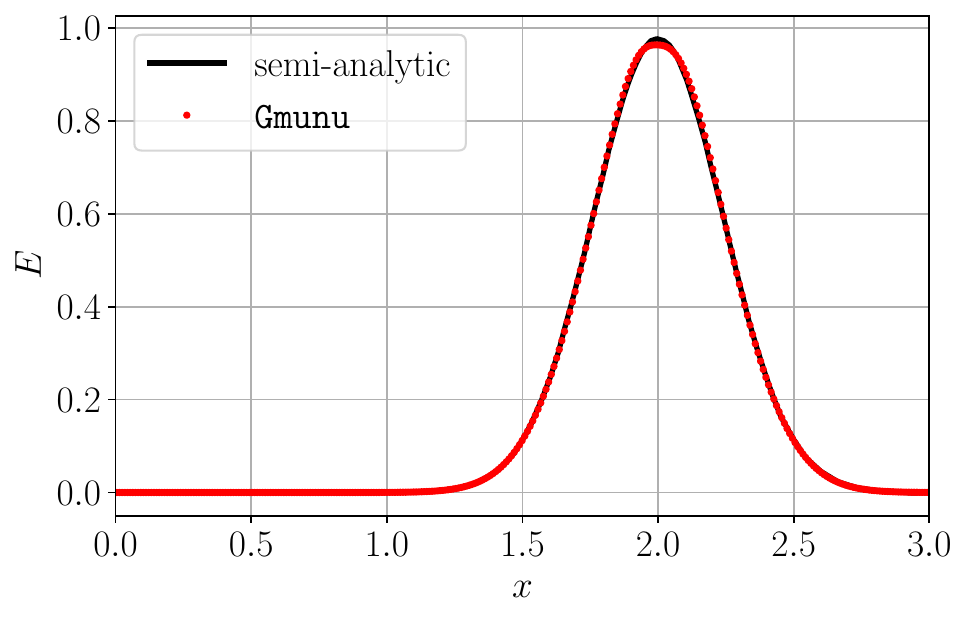}
	\caption{
		The energy density profile of the radiation ${E}$ at $t=4$ in a purely scattering medium with high scattering opacity $\bar{\kappa}_s = 10^3$ which moves with a relativistic velocity $v^x=0.5$.
		The red dots show the numerical solution obtained by \texttt{Gmunu} while the black solid line shows the corresponding semi-analytic solution.
		Note that this is known to be a demanding test, in the sense that the result is highly sensitive to the treatment of the optically thick limit \cite{2022MNRAS.512.1499R}.
		This figure shows that the results returned by \texttt{Gmunu} agree with the semi-analytic solution.
		}
	\label{fig:M1_diffusive_limit_moving_1d}
\end{figure}

Figure~\ref{fig:M1_diffusive_limit_moving_1d} shows the radiation energy density profile at time $t=4$.
As shown in the figure, the results obtained by using \texttt{Gmunu} agree with the semi-analytic reference solution.

\subsection{Diffusive point source}
Here, we present a test which also focuses on the scattering regime, i.e. the diffusive point source test proposed by \cite{2000MNRAS.317..550P}.
This test describes the evolution of a single point source in the diffusive limit.
The initial condition, and also the analytic solution, is given by
\begin{align}
	{E}\left(r,t\right) = &\left( \frac{\bar{\kappa}_s}{t}\right)^{N_{\text{dim}}/2} \exp\left( \frac{-3\bar{\kappa}_s r^2}{4ct}\right), \\
	{F}^r\left(r,t\right) = &\frac{r}{2t} {E}\left(r,t\right),
\end{align}
where $N_{\text{dim}}$ is the number of dimensions, which is set to be 2.
In this test, we consider a purely scattering medium with a scattering opacity $\bar{\kappa}_s = 100$.
The simulation starts from $t=1$ to $t=4$.
We assume cylindrical geometry in one dimensional, and the computational domain covers the region $\left[ 0, 1 \right]$ with 128 grid points. 

\begin{figure}
	\centering
	\includegraphics[width=\columnwidth, angle=0]{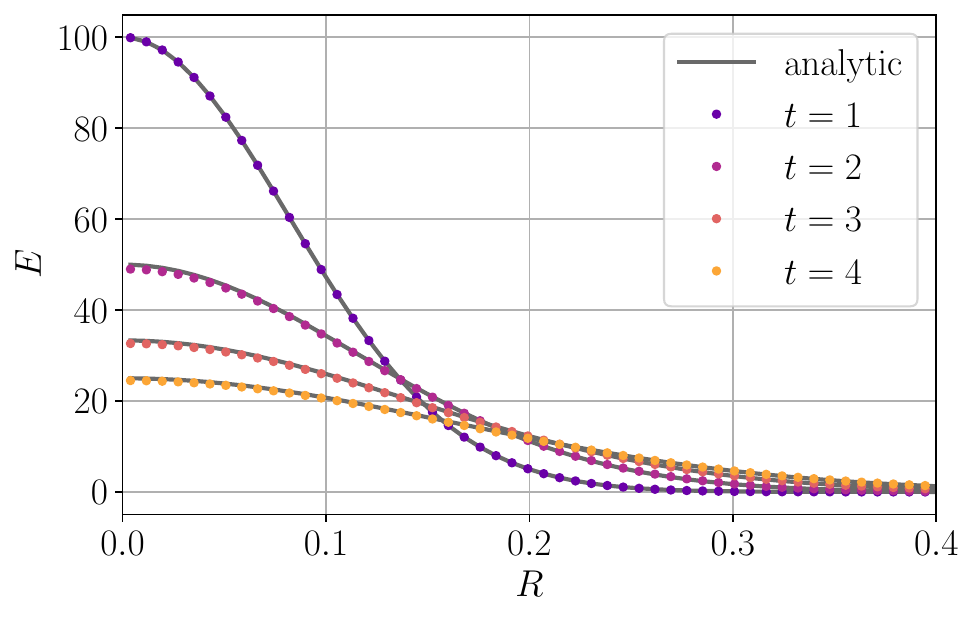}
	\caption{
		Comparison of the energy density profile $E$ at time $t=1,2,3,4$ to the analytic solution (grey solid lines).
		Since the simulation starts from $t=1$, the result is identical to the analytic solution.
		As shown in the figure, results produced by utilising \texttt{Gmunu} agrees with the analytic solution.
		}
	\label{fig:M1_diffusive_point_source_cylindrical_1d}	
\end{figure}

Figure~\ref{fig:M1_diffusive_point_source_cylindrical_1d} compares the energy density $E$ at different time $t$ to the analytic solution.
Since the simulation starts from $t=1$, the \texttt{Gmunu} result is identical to the analytic solution.
As shown in the figure, the \texttt{Gmunu} result agrees with the analytic solution.

\subsection{Shadow casting problems}
Here we present multidimensional tests which describe the interaction between radiation and a cylinder with high absorption opacity.

Firstly, we consider a radiation beam propagating from left to right.
The initial condition is given as
\begin{align}
	{E} =
	\begin{cases}
		1, 	&	\text{if } x \leq -0.4 \text{ and } y \in \left[-0.12,0.12\right];\\
		10^{-16}, 	&	\text{ otherwise},
	\end{cases}
\end{align}
and we set $\left( {F}_x/{E}, {F}_y/{E}, {F}_z/{E} \right) = \left( 0.999999, 0, 0 \right)$ everywhere in the computational domain.
In this test, we consider a cylinder of radius $R=0.07$ and located at $\left( -0.2,0\right)$ with a extremely high absorption coefficient $\bar{\kappa}_a = 10^6$.
This initial condition is kept fixed during the entire evolution for $x\leq -0.4$.
Note that this high absorption opacity $\bar{\kappa}_a$ is around six orders of magnitude larger than the radiation moments, resulting in significantly stiff interaction source terms in the evolution equations.
We choose this stiff situation on purpose to challenge the non-linear implicit solver and the IMEX time integrator implemented in \texttt{Gmunu}.
This test is run with a uniform grid $256 \times 128$ which covers the region $\left[-0.5,0.5\right]$ for $x$ and $\left[-0.25,0.25\right]$ for $y$.

\begin{figure}
	\centering
	\includegraphics[width=\columnwidth, angle=0]{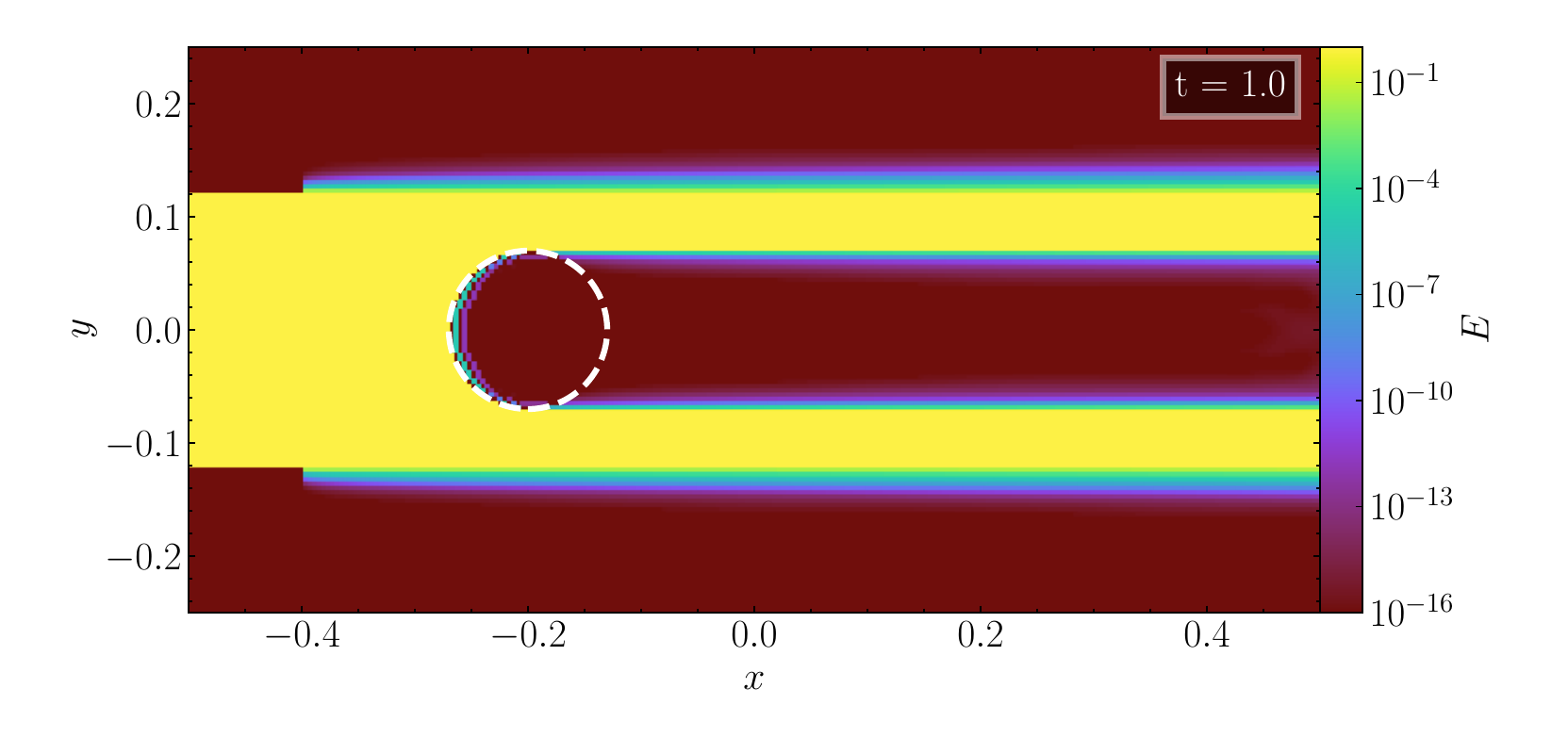}
	\caption{
		The radiation energy density profile at $t = 1$, where the initial conditions are kept fixed for $x \leq -0.4$ during the entire evolution.
		The radiation beam propagates from left to right, and is blocked by the optically thick cylinder (white dashed line) with high absorption opacity $\bar{\kappa}_a = 10^{6}$.
		As a result, this cylinder split the beam into two.
		The two beams keep propagating to the right, and a shadow is cast behind the cylinder.
		}
	\label{fig:M1_single_beam}	
\end{figure}

Figure~\ref{fig:M1_single_beam} shows the radiation energy density profile at $t = 1$.
As shown in the figure, the radiation beam propagates from left to right, and is obstructed by the optically thick cylinder.
This results in a shadow behind the cylinder, and split the beam into two which keep propagating to the right.

Shadow casting tests have been carried out in more non-trivial geometries by considering a point source \cite{2015MNRAS.453.3386J, 2016ApJS..222...20K, 2018ApJ...854...63O}.
Similar to \cite{2015MNRAS.453.3386J, 2016ApJS..222...20K, 2018ApJ...854...63O}, here we consider a point source which located at the origin $r=0$ with the radius $r_{\rm{src}}=1.5$.
The source has the absorption opacity
\begin{equation}
	\kappa_a = 10 \exp\left[ -\left(4 r / r_{\rm src}\right)^2 \right]
\end{equation}
and emissivity $\eta = \kappa_a J_{\text{eq}}$, where we choose $J_{\rm{eq}}=1$.
In addition, we also consider a purely absorbing sphere $(\bar{\eta}=0=\bar{\kappa}_s)$, with radius $r_{\rm shadow}=2$, located at a distance of $d=8$ from the centre of the source.
Unlike \cite{2015MNRAS.453.3386J, 2016ApJS..222...20K, 2018ApJ...854...63O}, here we consider a high absorption opacity $\kappa_a=10^6$ for the sphere.
To make the setup slightly asymmetric along the $x-y$ plane, instead of placing the absorbing sphere at the equatorial plane, we place it at the polar angle $\theta={11}\pi/{24}$ (i.e. the $z$ coordinate of the centre of the absorbing sphere is $d \cos(\theta)$).
This test is run in cylindrical coordinate $(R,z)$ with a uniform grid $128 \times 128$ which covers the region $\left[0,12\right]$ for $R$ and $\left[-6,6\right]$ for $z$.

\begin{figure}
	\centering
	\includegraphics[width=\columnwidth, angle=0]{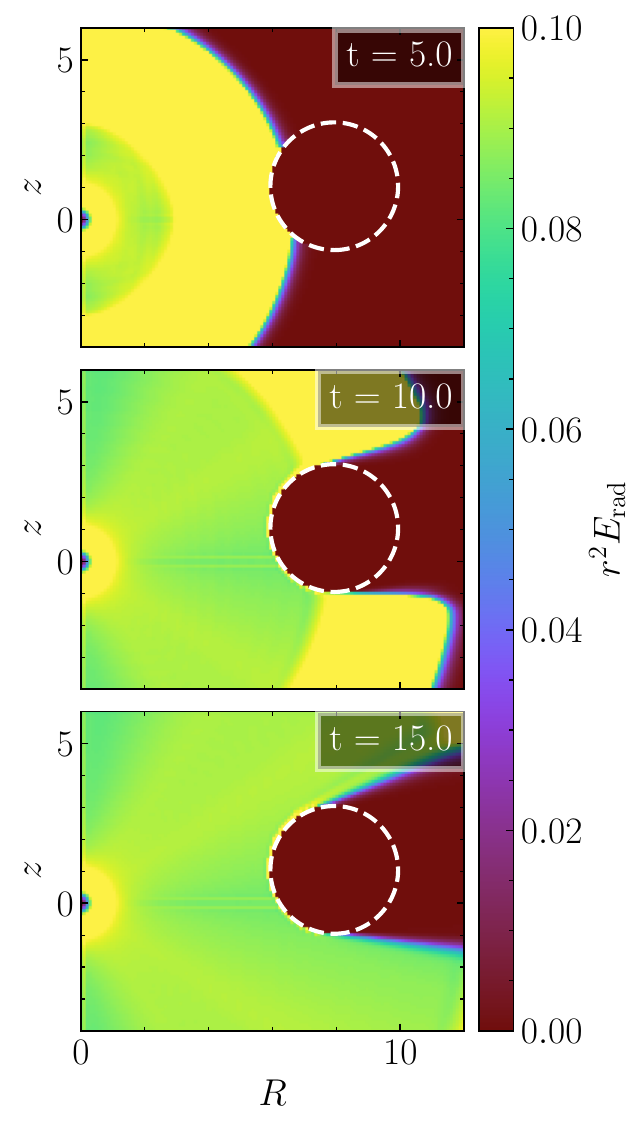}
	\caption{
		The radiation energy density profile scaled with $r^2$ at different time slides.
		The radiation propagates spherically from the source to outside, and is blocked by the optically thick sphere (white dashed line) with high absorption opacity $\bar{\kappa}_a = 10^{6}$.
		As in the single beam case (see figure~\ref{fig:M1_single_beam}), this sphere absorbs radiations and causes a shadow behind it.
		}
	\label{fig:M1_shadow_cylindrical_2d}	
\end{figure}

Figure~\ref{fig:M1_shadow_cylindrical_2d} shows the radiation energy density profile scaled with $r^2$ at $t = 5, 10, 15$, respectively.
The scaling of $r^2$ is to achieve a mostly constant value in the free-streaming regime (\cite{2018ApJ...854...63O}).
Similar to the single beam case discussed above (see figure~\ref{fig:M1_single_beam}), this sphere absorbs radiation and produces a shadow.
This indicates that our code behaves well in this test despite in 2D cylindrical coordinate with a off-angle setup.

\subsection{\label{sec:energy_adv_tests}Gravitational redshift and Doppler shift}
The tests presented above are all frequency-integrated tests.
To test if the code handles the energy-coupling terms for gravitational redshift and Doppler shift (see section~\ref{sec:energy_advection}) correctly, we perform the test introduced by \cite{2010ApJS..189..104M}, which has been used as a branch mark test by several authors, e.g. \cite{2015ApJS..219...24O, 2016ApJS..222...20K, 2020ApJ...900...71A, 2020MNRAS.496.2000C}.

To separately assess the handling of Doppler shift (with non-zero velocity profile), gravitational redshift (with curved spacetime) and the combination of these two, we mainly follow the setup in \cite{2010ApJS..189..104M}.
In particular, we consider a sphere with radius $R = 10 \text{ km}$ with a uniform density $\rho = 9 \times 10^{14} \text{ g/}\text{cm}^3$.
In addition, we consider a sharp velocity profile which mimics the accretion phase of core-collapse supernovae
\begin{align}
	v_r = 
	\begin{cases}
		0,	&	\text{if } r \leq 135 \text{ km} ;\\
		-0.2 c \left( \frac{ r - 135 \text{ km}}{ 150 \text{ km} - 135 \text{ km}} \right),	&	\text{if } 135 \text{ km} < r \leq 150 \text{ km} ;\\
		-0.2 c \left( \frac{ 150 \text{ km}}{ r} \right)^2,	&	\text{if } r > 150 \text{ km} .
	\end{cases}
\end{align}
Given the rest-mass density and velocity profiles, the metric quantities such as the conformal factor $\psi$, the lapse function $\alpha$ and the shift vector $\beta^i$ are calculated by utilising the metric solver in \texttt{Gmunu} \cite{2020CQGra..37n5015C,2021MNRAS.508.2279C}.
The initial neutrino profile is set as follows:
\begin{align}
	( \mathcal{E}, \mathcal{F}_r/\mathcal{E} )=
	\begin{cases}
		\left( \mathcal{B}, 10^{-2}\right)	&	\text{if } r \leq R;\\
		\left( \mathcal{B} \left(\frac{R}{r}\right)^2, 1-10^{-3}\right)	&	\text{if } r > R,
	\end{cases}
\end{align}
where $\mathcal{B} $ is the black body function, which is a function of frequency $\nu$, chemical potential $\mu$ and temperature $T$.
Specifically, under the chosen convention, the black body function is given by
\begin{equation}
	\mathcal{B} \left(\nu, \mu, T \right) = \frac{\nu}{\exp\left[ \left(h\nu - \mu\right)/{k_{\rm{B}}T}\right] + 1 },
\end{equation}
where $h$ is Planck constant and $k_{\rm{B}}$ is Boltzmann constant.
In this test, the chemical potential is chosen to be $\mu=0$ and the temperature is set to be $T = 5 \text{ MeV}$.
The absorption coefficient $\kappa_a$ is set to be 60 $\text{cm}^{-1}$ in the sphere ($r \leq R$) while vanishing elsewhere.
The emissivity is simply $\eta = \kappa_a \mathcal{B}$.
In this test, we consider spherical coordinate in 1D. 
The computational domain covers $[0,10^4] \text{ km}$ for $r$, with the resolution {$N_r = 128$} and allowing 6 mesh levels (an effective resolution of $4096$).
The refinement level is decided by a ratio $\Delta r / r$.
In particular, we refine the block if $\Delta r / r > 0.01$ in any of the grid in the block.
The frequency space is discretised into 18 groups logarithmically from $1 \text{ MeV} / h$ to $280 \text{ MeV} /h$.
The simulation is performed until the system reaches a stationary state; the results are extracted at $t_{\max} = 1 \text{ s}$.

As discussed in \cite{2010ApJS..189..104M}, stationary solution is available for this test.
In particular, the average neutrino energy $\langle \varepsilon \rangle$ and the redshift-corrected luminosity $L_{\text{rs}}$ obey the following relations:
\begin{align}
	W \alpha \left(1+v_r\right) \langle \varepsilon \rangle = &  \text{constant}, \\
	\frac{1+v_r}{1-v_r} L_{\text{rs}} = & \text{constant}, 
\end{align}
where $W$ is the Lorentz factor.
The average neutrino energy and the redshift-corrected luminosity are defined as
\begin{align}
	\langle \varepsilon \rangle &= \frac{ \int_0^\infty \mathcal{J} \dd{V_\nu} }{ \int_0^\infty \mathcal{J}/\nu \dd{V_\nu} }, \text{ and } \\
	L_{\text{rs}} &= 4 \pi r^2 \alpha^2 \psi^4 \int_0^\infty \mathcal{H} \dd{V_\nu},
\end{align}
respectively.
In the following, the analytic solution are computed based on the value at the surface of the sphere, i.e. $\langle \varepsilon \rangle \left(r=R\right)$ and $L_{\text{rs}} \left(r=R\right)$.
In addition, the numerical results are scaled by their corresponding value at the outer boundary of the computational domain, i.e. $\langle \varepsilon \rangle_{\infty} \equiv \langle \varepsilon \rangle \left(r=10^4 \text{ km}\right)$ and $L_{\text{rs},\infty} \equiv L_{\text{rs}} \left(r=10^4 \text{ km}\right)$.

\begin{figure*}
	\centering
	\includegraphics[width=\textwidth, angle=0]{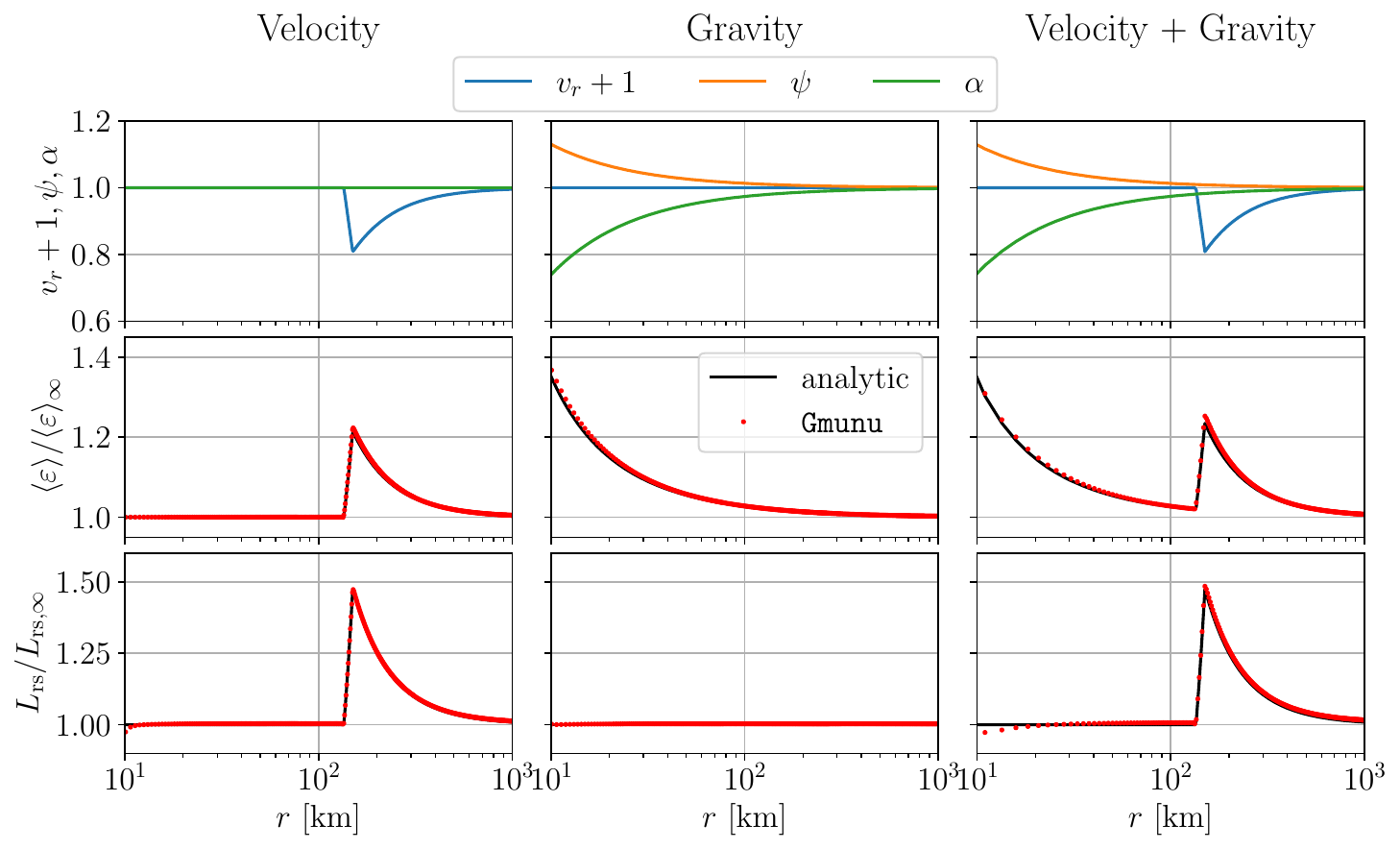}
	\caption{
	Comparison of the numerical solutions obtained by utilising \texttt{Gmunu} (red dots) and the analytic solutions (black solid lines) of the energy advection test introduced by \cite{2010ApJS..189..104M}.
	Three cases are considered in this test, namely, we consider
	(i) just the shape velocity profile without gravitational fields (\emph{left column});
	(ii) just the gravitational fields without velocity profile (\emph{middle column}), and 
	(iii) both the shape velocity profile and gravitational fields (\emph{right column}).
	The \emph{upper panels} show the shifted velocity profile $v_r + 1$ (blue solid lines), conformal factor $\psi$ (orange solid lines) and lapse function $\alpha$ (green solid lines) in different cases.
	The \emph{middle} and \emph{lower panel} compare of the simulated results of the rescaled average neutrino energy $\langle \varepsilon \rangle / \langle \varepsilon \rangle_{\infty}$ and the redshift-corrected luminosity $L_{\text{rs}} / L_{\text{rs},\infty}$ obtained by utilising \texttt{Gmunu} (red dots) and the analytic solutions (black solid lines).
	In all cases, results returned by \texttt{Gmunu} are in agreement with the analytic expressions.
	This test demonstrates that \texttt{Gmunu} is able to handle the frequency advection terms in the evolution of radiations, which corresponds to gravitational redshift and Doppler shift affects of the radiation with different frequencies.
		}
	\label{fig:M1_energy_advection_tests_1d}	
\end{figure*}

Figure~\ref{fig:M1_energy_advection_tests_1d} compares the rescaled average neutrino energy $\langle \varepsilon \rangle / \langle \varepsilon \rangle_{\infty}$ and the redshift-corrected luminosity $L_{\text{rs}} / L_{\text{rs},\infty} $ obtained by utilising \texttt{Gmunu} (red dots) and the analytic solutions (black solid lines).
As in \cite{2010ApJS..189..104M}, we consider three cases, namely, 
(i) the shape velocity profile in flat spacetime (left column);
(ii) vanishing velocities in curved spacetime (middle column), and 
(iii) the shape velocity profile in curved spacetime (right column).
The numerical results obtained by \texttt{Gmunu} agree with analytic results well in all three cases.
This test demonstrates that \texttt{Gmunu} is able to handle the advection terms in the frequency space, which corresponds to gravitational redshift and Doppler shift affects of the radiation with different frequencies.

\section{\label{sec:application_tests}Application examples}
The tests presented in the previous section considered only the monochromatic radiation source terms (i.e. there is no species or frequency couplings) and with idealised opacities.
To test if our code is able to handle the radiation where different species and frequencies are strongly coupled, in this section, we consider the neutrino transport in the context of core-collapse supernovae and hot neutron star as examples.

\subsection{\label{sec:nu_int}Neutrino source terms and couplings}
In addition to the source terms of emission, absorption, and elastic scattering, our application examples also incorporate the source terms of neutrino-lepton inelastic scattering, denoted as $\mathcal{S}^\mu_{\mathrm{IS}}$, and neutrino-pair processes, denoted as $\mathcal{S}^\mu_{\mathrm{Pair}}$. 
These terms describe the neutrino-electron inelastic scattering and electron-positron pair annihilation.
In this case, the radiation four-force for neutrinos can be written as \citep{2011PThPh.125.1255S, 2015ApJS..219...24O} 
\begin{equation}\label{eq:nu_source_terms}
	\mathcal{S}^\mu_{\mathrm{rad}} = \mathcal{S}^\mu_{\mathrm{E/A}} + \mathcal{S}^\mu_{\mathrm{ES}} + \mathcal{S}^\mu_{\mathrm{IS}} + \mathcal{S}^\mu_{\mathrm{Pair}},
\end{equation}
where, for each species, the inelastic scattering $\mathcal{S}^\mu_{\mathrm{IS}}$ and the neutrino-pair processes $\mathcal{S}^\mu_{\mathrm{Pair}}$ terms are defined as
\onecolumngrid
\begin{align}
	&\begin{aligned}\label{eq:ine_scattering_source}
	\mathcal{S}^\mu_{\text{IS}} \left( \nu \right) = \int \frac{\dd V_{\nu'}}{\nu'} 
		\Big\{ 
		& \left\{ \left[ \nu-\mathcal{J}\left( \nu \right) \right] u^\mu - \mathcal{H}^\mu\left( \nu \right) \right\} \mathcal{J}'\left(\nu'\right) R^{\rm{in}}_0\left(\nu,\nu'\right) \\
		& + \frac{\mathcal{H}^\mu\left( \nu' \right)}{3} \left\{ \left[ \nu-\mathcal{J}\left( \nu \right) \right] R^{\rm{in}}_1\left(\nu,\nu'\right) +  \mathcal{J}\left(\nu\right)R^{\rm{out}}_1\left(\nu,\nu'\right) \right\} \\
		& - \left\{ h_{\alpha\beta} \mathcal{H}^\alpha\left(\nu\right) \mathcal{H}^\beta\left(\nu'\right)u^\mu + \tilde{\mathcal{K}}^{\mu\alpha}\left(\nu\right) \mathcal{H}_\alpha\left(\nu'\right) \right\} \left[R^{\rm{in}}_1\left(\nu,\nu'\right)-R^{\rm{out}}_1\left(\nu,\nu'\right)\right] \\
		& - \left[ \mathcal{J}\left(\nu\right) u^\mu + \mathcal{H}^\mu\left(\nu\right) \right] \left[ \nu' - \mathcal{J}\left( \nu' \right) \right]R^{\rm{out}}_0\left(\nu,\nu'\right)
		\Big\}
	\end{aligned} ,
	\\
	&\begin{aligned}\label{eq:pair_source}
	\mathcal{S}^\mu_{\text{Pair}} \left( \nu \right) = \int \frac{\dd V_{\nu'}}{\nu'} 
		\Big\{ 
		& \left\{ \left[ \nu - \mathcal{J}\left( \nu \right) \right] u^\mu - \mathcal{H}^\mu\left( \nu \right) \right\} \left[ \nu' - \bar{\mathcal{J}}\left( \nu' \right) \right] R^{\rm{pro}}_0\left(\nu,\nu'\right) \\
		& - \frac{\bar{\mathcal{H}}^\mu\left( \nu' \right)}{3} \left\{ \left[ \nu - \mathcal{J}\left( \nu \right) \right] R^{\rm{pro}}_1\left(\nu,\nu'\right) + \mathcal{J}\left(\nu\right)R^{\rm{ann}}_1\left(\nu,\nu'\right) \right\} \\
		& + \left\{ h_{\alpha\beta} \mathcal{H}^\alpha\left(\nu\right) \bar{\mathcal{H}}^\beta\left(\nu'\right)u^\mu + \tilde{\mathcal{K}}\left(\nu\right)^{\mu\alpha} \bar{\mathcal{H}}_\alpha\left(\nu'\right) \right\} \left[R^{\rm{pro}}_1\left(\nu,\nu'\right)-R^{\rm{ann}}_1\left(\nu,\nu'\right)\right] \\
		& - \left[ \mathcal{J}\left(\nu\right) u^\mu + \mathcal{H}^\mu\left(\nu\right) \right] \bar{\mathcal{J}}\left( \nu' \right) R^{\rm{ann}}_0\left(\nu,\nu'\right)
		\Big\}
	\end{aligned} .
\end{align}
\twocolumngrid
\noindent Here, $\tilde{\mathcal{K}}^{\mu\alpha}\left(\nu\right) \equiv \mathcal{K}^{\mu\alpha}\left(\nu\right)-h^{\mu\alpha}\mathcal{J}\left(\nu\right) / 3$ is the traceless part of $\mathcal{K}^{\mu\alpha}\left(\nu\right)$.
Quantities with bar such as $\bar{\mathcal{J}}$ and $\bar{\mathcal{H}}^\mu$ denote the radiation moments for anti-neutrinos.
$R^{\rm{in}}$ and $R^{\rm{out}}$ are the kernels of inelastic scattering while $R^{\rm{pro}}$ and $R^{\rm{ann}}$ are the production and annihilation kernels of neutrino-pair processes.
As shown in equation~\eqref{eq:ine_scattering_source} and \eqref{eq:pair_source}, the radiation source term for each species and frequency involves not only the radiation at other frequency-bins, but also different species (its anti-particle).

Note that the computation of neutrino opacities and the kernels are non-trivial.
However, the discussion of which is beyond the scope of this work.
Currently, the neutrino opacities and kernels are provided by either tabulating \texttt{NuLib}\footnote{\texttt{NuLib} is an open-source library, available at \url{http://www.nulib.org}.} \citep{2015ApJS..219...24O} tables, or coupling to our newly developed neutrino microphysics library \texttt{Weakhub} \citep{2023_weakhub}.
To maintain consistent comparisons to the work in the literature, we consider the conventional set of interactions as in \cite{2005ApJ...620..840L, 2010ApJS..189..104M, 2015ApJS..219...24O, 2018JPhG...45j4001O} and ignoring the weak-magnetism and recoil corrections \citep{2015ApJS..219...24O}.
This set of interactions are summarised in table~\ref{tab:nu_interactions}.
Note that, by following \cite{2015ApJS..219...24O}, we approximate the treatment for neutrino-pair processes such as electron-positron annihilation and nucleon-nucleon Bremsstrahlung.
Specifically, the thermal processes for electron type neutrinos and anti-neutrinos are not included.
In addition, the full neutrino-matter interaction terms for heavy-lepton neutrinos are approximately represented with effective emissivity and absorption opacities.
As a result, there is no neutrino species coupling with this set of neutrino interaction.
Therefore, \emph{multi-species multi-group} implicit solver is not necessary with these neutrino interactions.

\begin{table}
\centering
\begin{tabular}{ll}
    \hline
    Beta processes                              & Neutrino-pair processes \\
    \hline
    $\nu_e+n \leftrightarrow p+e^{-}$           & $e^{-}+e^{+} \leftrightarrow \nu+\bar{\nu}$ \\
    $\bar{\nu}_e+p \leftrightarrow n+e^{+}$     & $N+N \leftrightarrow N+N+\nu+\bar{\nu}$ \\
    $\nu_{e}+(A,Z-1) \leftrightarrow (A,Z)+e^-$ & \\
     \hline
    Elastic scattering                          & Inelastic scattering \\
     \hline
    $\nu+N \leftrightarrow \nu+N$               & $\nu+e^{-}  \leftrightarrow \nu+e^{-}$\\
    $\nu+(A,Z) \leftrightarrow \nu+(A,Z)$       & \\
    $\nu+\alpha \leftrightarrow \nu+\alpha$     & \\
    \hline
    \end{tabular}
\caption{\label{tab:nu_interactions}
       Conventional set of neutrino interactions considered in this work.
       Here we denote the electron, anti-electron and heavy-lepton neutrino as $\nu_e$, $\bar{\nu}_e$ and $\nu_x$, respectively.
       $\nu$ represents all three species of neutrino.
       Interactions that involve a specific type of neutrino are expressed explicitly.
       $(A,Z)$ represents a heavy nucleus with a mass number of $A$ and a proton number of $Z$, without including $\alpha$ particle.
       The neutrino-pair processes could be either approximately treated as effective emissivity/absorption opacity or handled by using the full production/annihilation kernels.
       }
\end{table}
The evolution of the electron fraction $Y_e$, which is defined as the number of electrons per baryon, has to be included in order to describe matter in nuclear statistical equilibrium and compute the neutrino emissivity/opacities.
Since the $\beta$-processes of $\nu_e$ and $\bar{\nu}_e$ change the electron fraction $Y_e$, the source term of the evolution equation of the electron fraction $Y_e$ (see \cite{2023_leakage}) is expressed as
\begin{equation}\label{eq:evolution_ye}
	s_{D Y_e} = m_{\rm{u}} \int \frac{\dd V_{\nu'}}{\nu'} \left[ {s}_{{\rm rad},\nu_e}^\mu\left(\nu'\right) - {s}_{{\rm rad},\bar{\nu}_e}^\mu\left(\nu'\right) \right] u_\mu , 
\end{equation} 
where $m_{\rm{u}}$ is the atomic mass unit.
This coupling is treated explicitly as discussed in section~\ref{sec:coupling}.
Specifically, the conserved quantity for electron fraction $q_{D Y_e}$ is updated by
\begin{equation} 
\begin{aligned}
        & q_{D Y_e} \rightarrow q_{D Y_e} \\
		& + \Delta t \left\{ m_{\rm{u}} \int \frac{\dd V_{\nu'}}{\nu'} \left[ {s}_{\rm{rad},\nu_e}^\mu\left(\nu'\right) - {s}_{\rm{rad},\bar{\nu}_e}^\mu\left(\nu'\right) \right] u_\mu \right\}
\end{aligned}
\end{equation} 
once the radiation moments are solved implicitly. 

\subsection{\label{sec:ccsn_test}Core collapse of a 15 $\rm{M_{\odot}}$ star in one dimension}
The collapse, bounce and early post-bounce evolution of the 15 $\rm{M_{\odot}}$ progenitor star s15s7b2 of \cite{1995ApJS..101..181W} has become a standard test for core-collapse supernovae simulation code (e.g. \cite{2005ApJ...620..840L, 2010ApJS..189..104M, 2015ApJS..219...24O, 2016ApJS..222...20K}).
In these work, the equation of state of \cite{1991NuPhA.535..331L} with an incompressibility parameter of $K=180 {\rm \ MeV}$ is used.
Note that this equation of state, which has a maximum cold neutron star gravitational mass of $1.84 {\rm \ M_{\odot}}$, has been ruled out already.
For the purposes of this comparison, we use the same equation of state.

In this section, we present the core-collapse supernovae simulation with the same progenitor, and compare our result with the one of \texttt{AGILE-BOLTZTRAN}, \texttt{VERTEX} and \texttt{GR1D} \citep{2010CQGra..27k4103O, 2015ApJS..219...24O}.
The data of \texttt{AGILE-BOLTZTRAN} and \texttt{VERTEX} are obtained from the online material provided in the electronic version of \cite{2005ApJ...620..840L} while the data of \texttt{GR1D} is reproduced by using the code with the settings for the section~5.1 in \cite{2015ApJS..219...24O}\footnote{\texttt{GR1D} is an open-source neutrino radiation transport code for core-collapse supernovae \citep{2010CQGra..27k4103O, 2015ApJS..219...24O}.
{
The \texttt{GR1D} code, and also the parameter files, equation of state and the \texttt{NuLib} tables used in \cite{2015ApJS..219...24O}, are available at \url{http://www.GR1Dcode.org}.
}
Note that, since the conformally flat metric equations are equivalent to the Einstein equations in spherical symmetry, our results here are fully general relativistic as in \texttt{GR1D} and \texttt{AGILE-BOLTZTRAN}.
}.
In this subsection, we use the identical \texttt{NuLib} table.

\subsubsection{\label{sec:ccsn_imp_mode}Treatments in different phases}
As mentioned in section~\ref{sec:rank}, avoiding full implicit treatment would significantly reduce the computational cost.
In fact, in the context of core-collapse supernovae, given that the timestep is properly chosen, a full implicit treatment which includes fluid variables is barely necessary even when the system is stiff \citep{2015ApJS..219...24O, 2015MNRAS.453.3386J, 2020LRCA....6....4M}.
For instance, in optically thick region, neutrinos are trapped in the fluid and are very close to weak equilibrium.
The net change (absorption minus emission) of the frequency-integrated neutrinos source terms are effectively small.
As a result, the change on fluid quantities due to neutrinos are negligible compared with hydrodynamical effects.
In addition to the fluid quantities, depending on the stage of the collapse, it is also valid to treat part of the neutrino source terms explicitly, which could significantly reduce the size of the non-linear system.

In practice, we split the simulation into three phases.
In phase 1, the collapse begins but not extremely dynamical.
We update the metric at every $0.1 {\rm \ ms}$, set the Courant–Friedrichs–Lewy (CFL) factor to be 0.8, and check the refinement criteria at every 10 iterations.
In this phase, we use mode 3 (\emph{single-species single-group}) radiation-interaction terms treatment (see section~\ref{sec:rank}).
Once the maximum rest mass density $\rho$ is larger than $10^{12} {\rm \ g/cm^3}$, we switch to phase 2, where we update the metric at every $0.01 {\rm \ ms}$, set the CFL factor to be 0.4, and check the refinement criteria at every iteration.
In this phase, we use mode 2 (\emph{single-species multi-group}) radiation-interaction terms treatment. 
Core bounce is expected in this phase, which is defined as when the matter entropy per baryon is larger or equals to 3 (i.e. $s \geq 3 \ k_{\rm{B}} / {\rm baryon}$) in the core region.
We monitor this core-bounce criteria in the core region (i.e. $r \lesssim 30 {\rm \ km}$) at each timestep in this phase.
Finally, we switch to phase 3 (post-bounce phase) 20 ms after core bounce.
In this phase, the treatment for radiation-interaction terms is unchanged.
We update the metric at every $0.05{\rm \ ms}$, set the CFL factor to be 0.6, and check the refinement criteria at every 5 iterations.

Since the electron fraction $Y_e$ is not solved consistently in the implicit step, it is possible that the change of the electron fraction is too large, resulting non-physical result and eventually crash the code.
Similar in \cite{2015ApJS..219...24O, 2015PhRvD..91l4021F}, we monitor the change of the electron fraction $Y_e$ at each time step.
When the relative difference of the electron fraction $Y_e$ is larger than $10^{-3}$, we scale down the CFL factor by multiplying by 0.9 and continue the simulation.
Otherwise, we scale up the Courant–Friedrichs–Lewy factor by dividing by $0.9$ until it goes back to the corresponding setting in the particular phase of the simulation.

\subsubsection{Numerical setup}
The computational domain covers $[0,10^4] \text{ km}$ for $r$, with the resolution {$N_r = 128$} and allowing $l_{\max}=12$ mesh levels.
For the refinement criteria, we apply the L{\"o}hner's error estimator \citep{1987CMAME..61..323L,2021MNRAS.508.2279C} on the logarithmic rest mass density $\log_{10}\left( \rho \right)$.
This can effectively capture the sudden change of rest mass density (usually arise at shock) while keeping the refinement low elsewhere.
On top of the error estimator, to better resolve the high density region of the star, we require the block to the finest level $l_{\max}$ when any of the rest mass density $\rho$ in this block is larger than $\rho_{\rm{thr}} \equiv 5 \times 10^{12} { \rm \ g/cm^3}$.
Since we are mainly interested in the inner part of the massive star in this work, we further impose a maximum allowed refinement level at different location.
For instance, when the smallest radius $r_{\min}$ in a block is smaller than $100 {\rm \ km}$ (i.e. when $r_{\min} \leq 100 {\rm \ km}$), the highest allowed refinement level is $l_{\max}$.
Also, when $r_{\min} \leq 2 \times 100 {\rm \ km}$, the highest allowed refinement level is $l_{\max}-1$, so on and so forth.
We also enforce the refinement level to be lowest when the block contains outer boundaries.

The frequency space is discretised into 18 groups logarithmically from $1 \text{ MeV} / h$ to $280 \text{ MeV} /h$.
We evolve 3 species of neutrinos, namely, the electron neutrino $\nu_e$, anti-electron neutrino $\bar{\nu}_e$ and heavy-lepton neutrino $\nu_x$, where the muon and tauon neutrinos (i.e. $\nu_{\mu}, \bar{\nu}_{\mu}, \nu_{\tau}$ and $\bar{\nu}_{\tau}$) are grouped into $\nu_x$.

\subsubsection{Results}
\begin{figure}
	\centering
	\includegraphics[width=\columnwidth, angle=0]{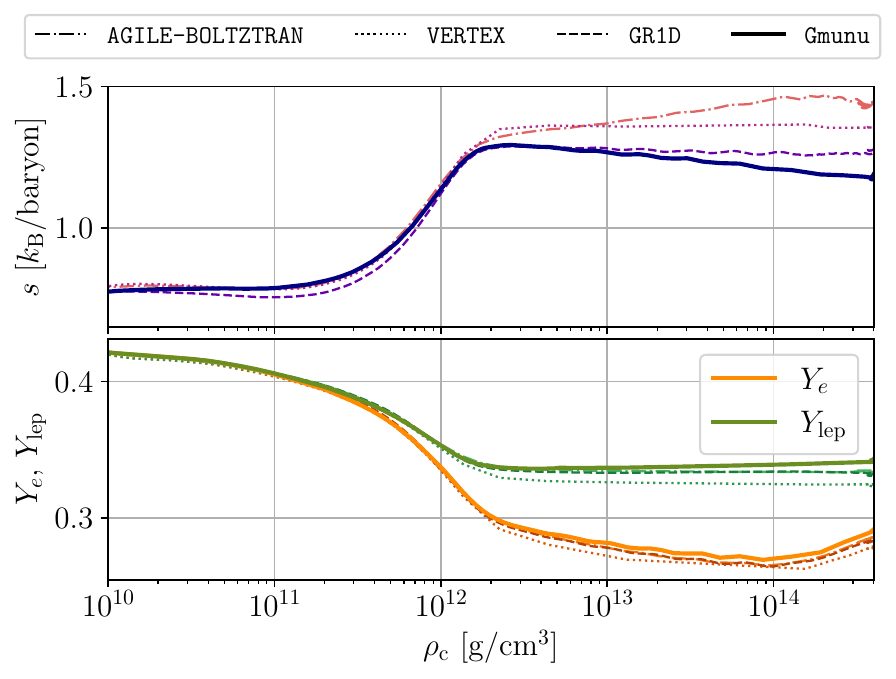}
	\caption{
	Evolution of central matter entropy per baryon ($s$, \emph{upper panel}), electron and total lepton number fractions ($Y_e$ and $Y_{\rm{lep}} \equiv Y_e + Y_\nu$, \emph{lower panel}) as functions of central density $\rho_c$ of a collapsing 15 ${\rm M_{\odot}}$ star before core-bounce.
	The solid lines show the numerical results obtained by \texttt{Gmunu}. 
	Note that the evolution of lepton numbers is highly sensitive to the implementations of the multi-group coupling and radiation advection in optically thick regions.
	Our results agree very well with the results of \texttt{AGILE-BOLTZTRAN} (dashed-dotted lines), \texttt{VERTEX} (dotted lines) and \texttt{GR1D} (dashed lines).
		}
	\label{fig_M1_ccsn_s15_1d_compare_center}	
\end{figure}

Figure~\ref{fig_M1_ccsn_s15_1d_compare_center} shows the evolution of central matter entropy per baryon $s$, central electron fraction $Y_e$ and lepton number fraction $Y_{\rm{lep}} \equiv Y_e + Y_\nu$ as functions of central density $\rho_c$ of the collapsing 15 $\rm{M_{\odot}}$ star before core-bounce.
During the deleptonization phase, the entropy per baryon increase due to neutrino interaction.
The core deleptonization ends when the central density reaches approximately $2\times 10^{12} {\rm \ g/cm^3}$.
Since then, the neutrinos are mostly trapped, where the lepton number fraction remain almost unchanged.
In this stage, the inner core collapses adiabatically, and the entropy per baryon remains nearly constant.
It is worthwhile to point out that the evolution of lepton numbers is highly sensitive to the implementations of the multi-group coupling, radiation space advection in optically thick regions and advection in frequency space for lepton number conservation, even when the exact same neutrino microphysics is used \citep{2015ApJS..219...24O, 2016ApJS..222...20K}.

\begin{figure*}
	\centering
	\includegraphics[width=\textwidth, angle=0]{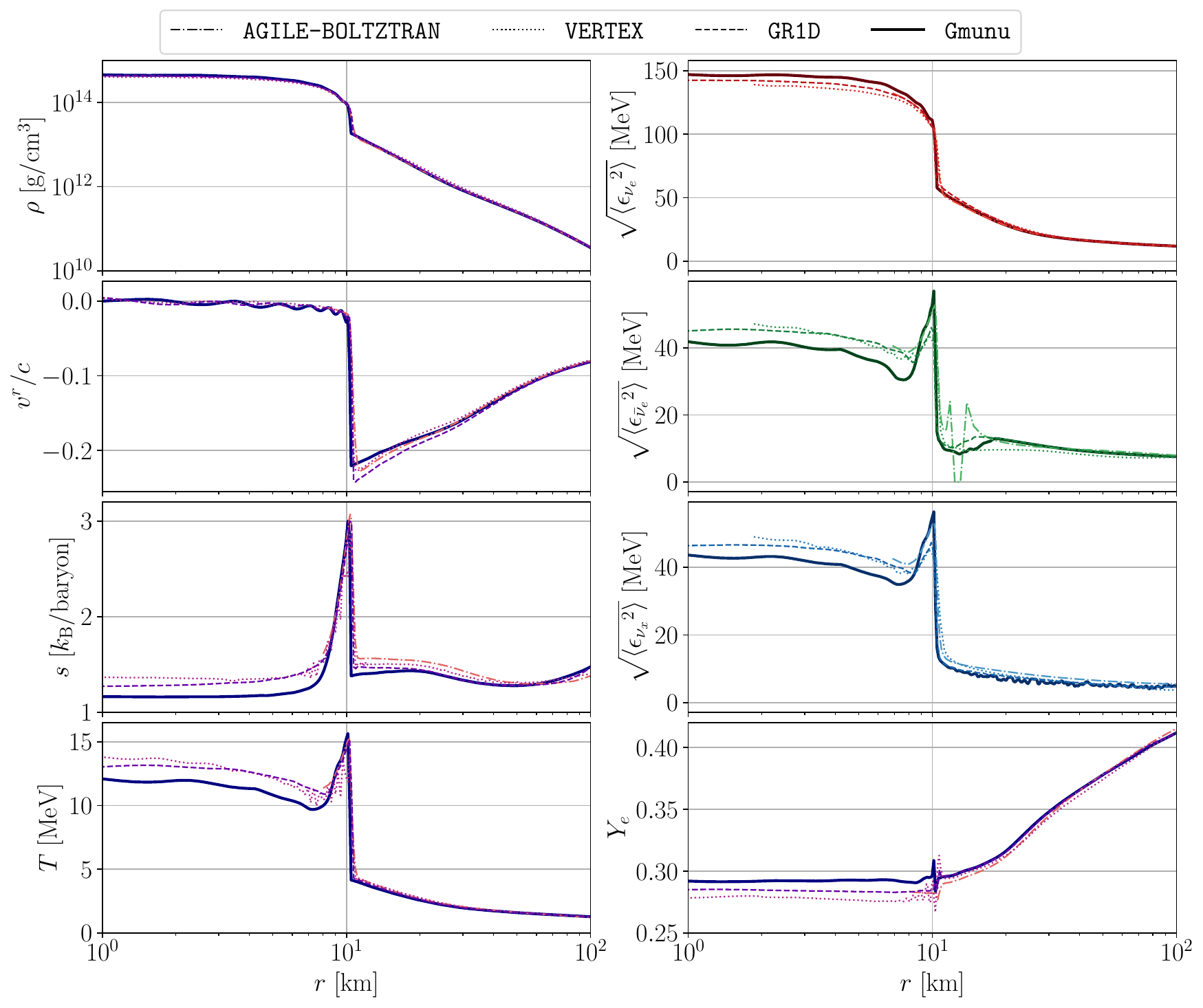}
	\caption{
		Comparison of the radial profiles of several quantities between {our code \texttt{Gmunu} and the reference codes (\texttt{AGILE-BOLTZTRAN}, \texttt{VERTEX} and \texttt{GR1D})} at the moment of core bounce.
		For instance, we compare rest mass density $\rho$, radial velocity $v^r/c$, matter temperature $T$, matter entropy per baryon $s$, neutrino root mean squared energy $\sqrt{\langle\epsilon_{\nu_i}^2\rangle}$ and electron fraction $Y_e$.
		The solid lines show the numerical results obtained by \texttt{Gmunu}.
		A shock is formed at $\sim 10 {\rm\ km}$, and the matter entropy reaches $3 \ k_{\rm{B}} / {\rm baryon}$ at the shock.
		Our results generated by \texttt{Gmunu} are quantitatively agreeing with the reference solution produced by \texttt{AGILE-BOLTZTRAN} (dashed-dotted lines), \texttt{VERTEX} (dotted lines) and \texttt{GR1D} (dashed lines).
		}
	\label{fig_M1_ccsn_s15_1d_compare_hydro}	
\end{figure*}

Strong repulsive forces of nuclear matter arise when the rest mass density exceeds nuclear saturation densities.
This results in core bounce, and forms the bounce shock.
In our simulation, the core bounces at $t_{\text{bounce}} \approx 178.21$ ms.
In figure~\ref{fig_M1_ccsn_s15_1d_compare_hydro}, we compare the radial profiles of several quantities among different codes at the moment of core bounce.
In particular, we compare rest mass density $\rho$, radial velocity $v^r/c$, matter temperature $T$, matter entropy per baryon $s$, neutrino root mean squared energy $\sqrt{\langle\epsilon_{\nu_i}^2\rangle}$ and electron fraction $Y_e$.
Here, the neutrino root mean squared energy is defined by
\begin{align}
	\sqrt{\langle\epsilon^2\rangle} \equiv \frac{ \int_0^\infty \nu \mathcal{J} \dd{V_\nu} }{ \int_0^\infty \mathcal{J}/\nu \dd{V_\nu} }.
\end{align}
As shown in the figure, a shock is formed at $\sim 10 {\rm \ km}$, and the matter entropy reaches $3 \ k_{\rm{B}} / {\rm baryon}$ at the shock.
Since the gauge adopted in \texttt{Gmunu} is different from the one in \texttt{AGILE-BOLTZTRAN}, \texttt{VERTEX} and \texttt{GR1D}, transformation is needed in order to compare the results directly.
The areal circumferential radius $r_{\rm{circ}}$ used in \cite{2005ApJ...620..840L} can be expressed in terms of the isotropic radial coordinate $r_{\rm{iso}}$ and the conformal factor $\psi$ by \citep{2006A&A...445..273M, 2010ApJS..189..104M}:
\begin{equation}
	r_{\rm{circ}} = \psi^2 r_{\rm{iso}}.
\end{equation}
Below, we simply use $r$ to denote the areal circumferential radius $r_{\rm{circ}}$.

{
Although the results produced by utilising \texttt{Gmunu} are quantitatively in good agreement with the reference models by using \texttt{AGILE-BOLTZTRAN}, \texttt{VERTEX} and \texttt{GR1D}, there are some deviations in the inner part of the star.
In particular, the hydrodynamical quantities such as entropy $s$, electron fraction $Y_e$ and temperature $T$ are slightly deviated from the reference solutions when the central rest mass density $\rho_c$ goes beyond $\sim 2 \times 10^{12} { \rm \ g/cm^3}$ (see figure~\ref{fig_M1_ccsn_s15_1d_compare_center}) and for the region where $r \lesssim 10~{\rm km}$ (see figure~\ref{fig_M1_ccsn_s15_1d_compare_hydro}).
These deviations could be due to the following three reasons.
Firstly, different implementations of nuclear equation of state (see also the discussion in \cite{2015ApJS..219...24O}) and the primitive recovery are very likely to cause the differences of the entropy and so as other hydrodynamical quantities.
Secondly, although both our code and \texttt{GR1D} couple the hydrodynamical quantities directly in the implicit radiation moment solver, such coupling will be applied twice for each timestep when IMEX-SSP2(2,2,2) is used.
The discrepancy caused by the direct coupling might be accumulated faster then in first-order implicit-explicit method.
Solving the entire evolution system including the hydrodynamical quantities implicitly will lead to more accurate and consistent results, which is left as future work.
Thirdly, since adaptive mesh refinement is used in our simulations, the resolution at the centre part is not fixed during the simulation, and depends on the rest mass density $\rho$ which changes rapidly right before core bounce.
The numerical errors due to such rapid refinements is one of the source of the error.
Refinement strategies that have better balance between the accuracy and computational cost will be explored in the future.
}

\begin{figure*}
	\centering
	\includegraphics[width=\textwidth, angle=0]{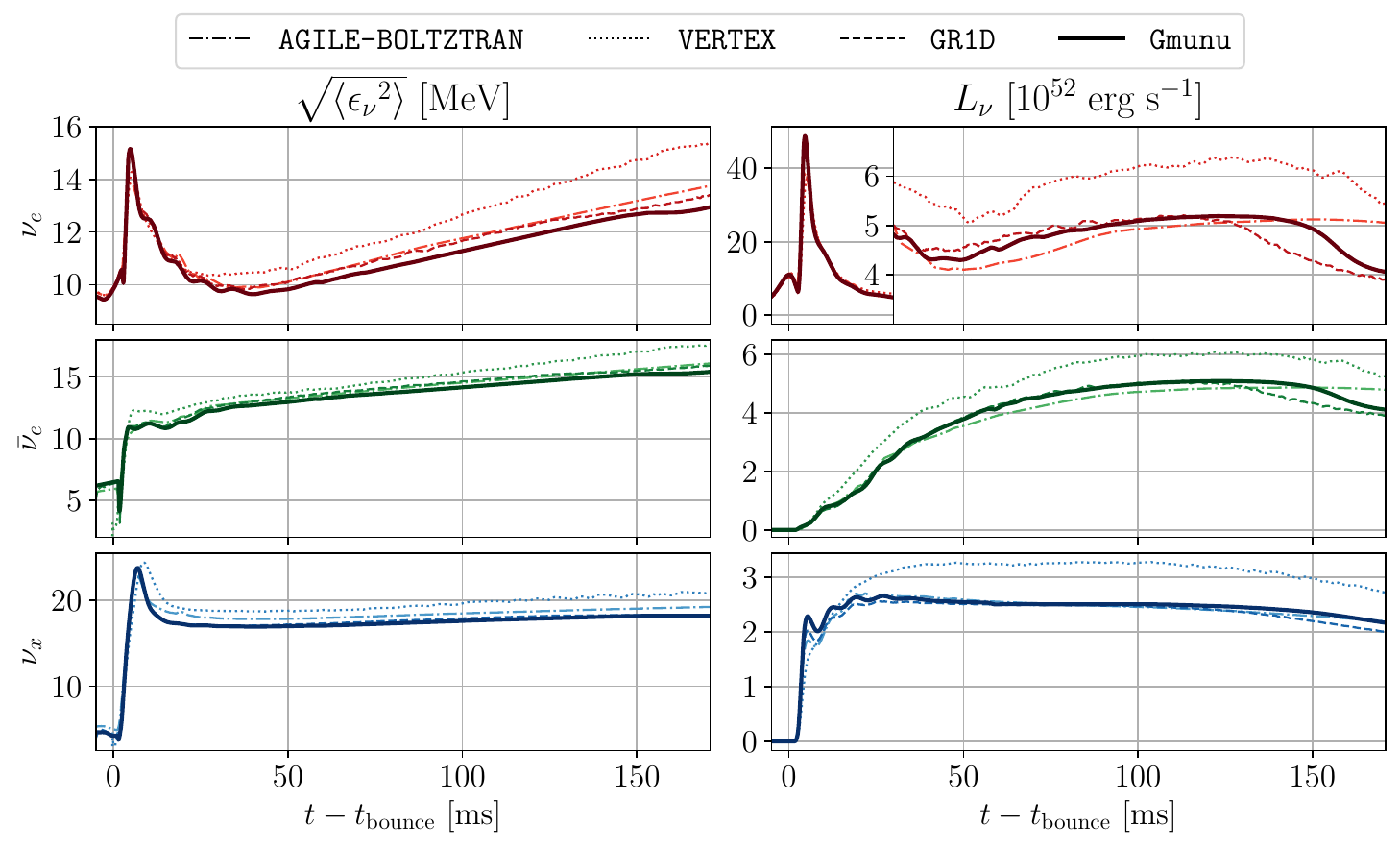}
	\caption{
	Time evolution of far-field neutrino root mean squared energies ($\sqrt{ \langle {\epsilon_\nu}^2 \rangle }$, \emph{left panel}) and luminosities ($L_\nu$, \emph{right panel}) measured by an observer comoving with fluid at 500 km of a collapsing 15 ${\rm M_{\odot}}$ star.
	The solid lines show the numerical results obtained by \texttt{Gmunu}.
	To compare the deleptonization burst and the early post-bounce evolution of the luminosity of electron type neutrino more in detail, we changed the scale after $t>30 {\rm{\ ms}}$ (\emph{upper right panel}).
	The evolution of these neutrino quantities are again highly sensitive to the implementation and essential to the predictions of observational signatures.
	Our results agree very well with the reference results produced by \texttt{AGILE-BOLTZTRAN} (dashed-dotted lines), \texttt{VERTEX} (dotted lines) and \texttt{GR1D} (dashed lines).
		}
	\label{fig_M1_ccsn_s15_1d_compare_nu_t}	
\end{figure*}

The far-field neutrino root mean squared energies and luminosities are important to the predictions of observation, which are highly sensitive to the microphysics considered and the implementation.
Therefore, it is necessary to show and compare these key neutrino quantities among codes.
In figure~\ref{fig_M1_ccsn_s15_1d_compare_nu_t} we show the time evolution of far-field neutrino root mean squared energies $\sqrt{ \langle {\epsilon_\nu}^2 \rangle }$ and luminosities $L_\nu$ measured by an observer comoving with fluid at 500 km.
As shown in the figure, the agreement between \texttt{Gmunu}, \texttt{AGILE-BOLTZTRAN} and \texttt{GR1D} is exceptional.
This is expected since both of two simulations adopt two-moment schemes and use identical neutrino opacities and kernels table.

\begin{figure}
	\centering
	\includegraphics[width=\columnwidth, angle=0]{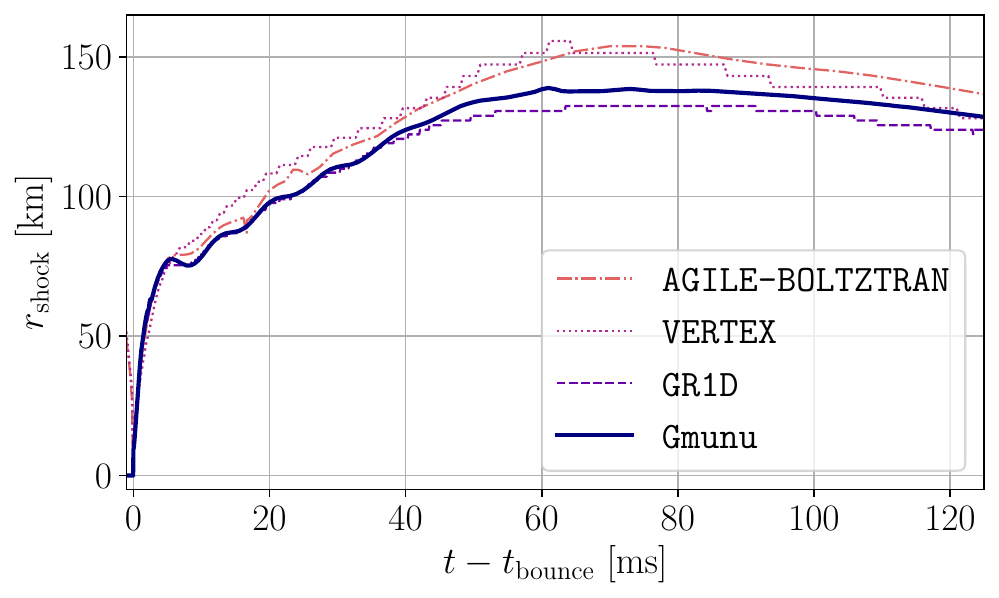}
	\caption{
		Time evolution of the shock radius.
		The solid lines show the numerical results obtained by \texttt{Gmunu}.
		The reference results produced by \texttt{AGILE-BOLTZTRAN}, \texttt{VERTEX} and \texttt{GR1D} are shown with dashed-dotted line, dotted line and dashed line, respectively.
		The shock radius predicted by \texttt{Gmunu} is mostly agreeing with \texttt{GR1D} for $t-t_{\rm{bounce}} \lesssim 40 {\rm \ ms}$ while it is approximately 10 km larger since then.
		Since then, it goes between the result of \texttt{AGILE-BOLTZTRAN} and \texttt{GR1D}.
		}
	\label{fig_M1_ccsn_s15_1d_compare_r_sh_t}	
\end{figure}

The shock radius evolution is also important in the core-collapse supernovae context.
Figure~\ref{fig_M1_ccsn_s15_1d_compare_r_sh_t} shows the time evolution of the shock radius.
Our results agree very well with \texttt{GR1D} for $t-t_{\rm{bounce}} \lesssim 40 {\rm \ ms}$.
After that, the shock radius predicted by \texttt{Gmunu} is roughly 10 km larger then \texttt{GR1D}'s, which lies between \texttt{AGILE-BOLTZTRAN} and \texttt{GR1D}.

The results presented above, especially the neutrino signals, are relatively closer to \texttt{GR1D}'s among the codes we have compared.
Despite the neutrino interactions considered are mostly the same in these runs, the way of how the neutrino opacities and kernels are computed could be different (e.g. the resolutions in energy space, number of species evolved are different).
Moreover, as mentioned above, approximated treatments for neutrino-pair processes are adopted in the \texttt{NuLib} table in this test.
Due to the fact that the same \texttt{NuLib} table with the same resolution in energy (frequency) space is used in both \texttt{GR1D} and \texttt{Gmunu}, the corresponding results are expected to be very similar.


\subsection{Hot neutron star}
For the second application example, we study the radial oscillation and neutrino emissions of a hot neutron star by following \cite{2013PhRvD..88f4009G, 2014PhRvD..89j4029N}.
In particular, we consider a non-rotating equilibrium model with SHT equation of state \cite{2011PhRvC..83c5802S} with central rest-mass density $\rho_c = 9.3 \times 10^{14} {\rm ~ g / cm^{3}}$.
The star has a constant entropy per baryon $s = 1 ~ k_{\rm{B}} / \text{baryon}$ and in $\beta$ equilibrium.
The temperature of which is roughly 30~MeV at the centre of the star.
The gravitational mass and the circumferential radius of this neutron star are $M_{\text{grav}} = 2.741 \text{ M}_{\odot}$ and $R_{\text{circ}} = 14 \text{ km}$, respectively.
{
As discussed in \cite{2013PhRvD..88f4009G}, this neutron star is mostly opaque to neutrinos, and the neutrino diffusion timescale ($\mathcal{O}(\rm{s})$) is much longer than its dynamical timescale ($\mathcal{O}(\rm{ms})$).
The neutron star is expected to be cooling slowly.
In addition, since the emitted neutrino will mostly be reabsorbed in the hot and dense region, the neutrino emission from the system are mostly come from the outer layer of the neutron star.
Therefore, this test problem, especially the neutrino emissions, is highly sensitive to the low density/atmosphere treatment.
}

The initial neutron star models are generated with the modified version of the open-source code \texttt{XNS}\footnote{{available at \url{https://www.arcetri.inaf.it/science/ahead/XNS/index.html}.}} \citet{2011A&A...528A.101B,XNS1,XNS2,XNS3}.
We simulate this initial model in one-dimensional spherical coordinates, where the computational domain covers $0 \leq r \leq 400 \;( \approx 591 \mathrm{ km})$ with the resolution {$N_r = 256$} and allowing 4 {refinement} levels (i.e., an effective resolution of {$N_r=2048$}).
For the simulations of neutron stars, we used the same refinement setting as in our previous work \cite{2021MNRAS.508.2279C}.
In particular, we defined a relativistic gravitational potential $\Phi \equiv 1 - \alpha$.
For any $\Phi$ larger than the maximum potential $\Phi_{\text{max}}$ (which is set as 0.2 in this work), the block is set to be finest.
While for the second-finest level, the same check is performed with a new maximum potential which is half of the previous one, so on and so forth.
The grid is fixed after the initialisation.
In this test, 2-nd order Montonized central (MC) limiter \cite{1974JCoPh..14..361V} is used. 
{
The rest mass density of the atmosphere $\rho_{\rm atmo}$ is set to be $10^{3} {\rm ~ g / cm^{3}}$.
}
The spacetime is kept fixed during the entire simulation (i.e. we evolve this system with Cowling approximation).

While different neutrino microphysics inputs are expected to affect the neutrino signals, the neutrino interactions we can include are so far limited by \texttt{NuLib}.
Instead of considering the same set of neutrino interactions as in \cite{2013PhRvD..88f4009G, 2014PhRvD..89j4029N}, we adopt essentially the same set of interactions as described in section~\ref{sec:nu_int}, except that the thermal processes for electron type neutrinos and anti-neutrinos (i.e. $e^- + e^+ \to \nu_e + \bar{\nu}_e$) are also included.
The inclusion of this interaction is to additionally test our \emph{multi-species multi-group} implicit solver, since the coupling of electron type neutrinos $\nu_e$ and anti-neutrinos $\bar{\nu}_e$ has to be taken into account.
{
Although this test problem is expected to be sensitive to the low density/atmosphere treatment, to achieve a stable evolution, the neutrino opacities and kernels are used only when the rest mass density $\rho$ is larger than $10^{11} {\rm ~ g / cm^{3}}$, which is eight orders of magnitude larger than the atmosphere density $\rho_{\rm atmo}$.
}
The neutrino moments are initialised by evolving the radiation sector while keeping the hydrodynamical profile fixed until the system reach to equilibrium.
In this test, we evolve the neutrinos while keeping the hydrodynamical profile fixed for 5 ms before the dynamical simulation.

\begin{figure}
	\centering
	\includegraphics[width=\columnwidth, angle=0]{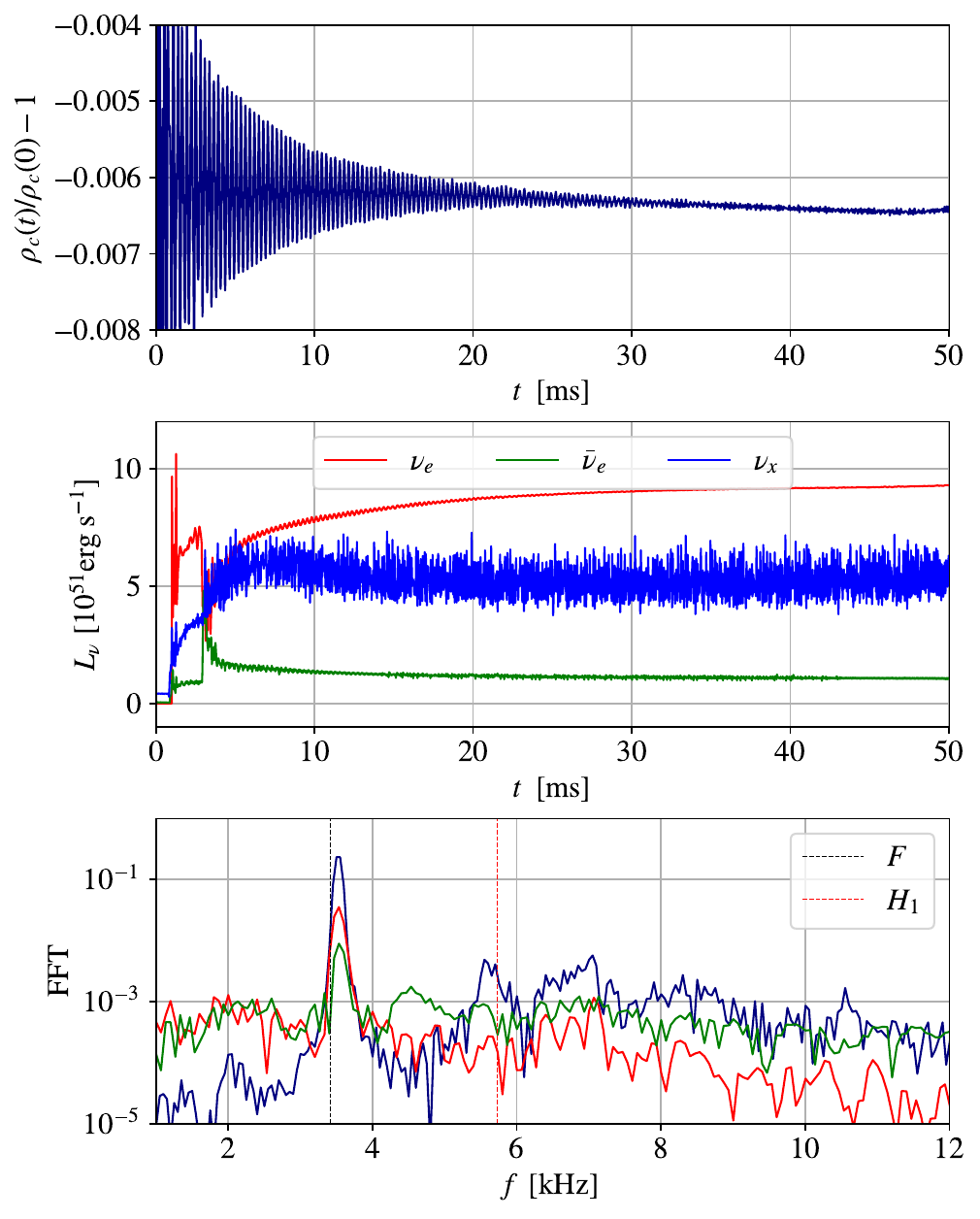}
	\caption{
                \emph{Upper panel}: The relative variation of the central rest mass density $\rho_c$ of the hot neutron star in time.
		{
		\emph{Middle panel}: Time evolution of far-field neutrino luminosities measured by an observer comoving with fluid at 100 km of a hot neutron star.
		The luminosities are zero for $t \lesssim 3$ ms as the neutrino signals are not yet arrived at the extraction point.
		}
		\emph{Lower panel}: The fast Fourier transform of the central rest mass density{
		, and the luminosities of electron and anti-electron neutrinos.
		The amplitude of the fast Fourier transform of the central rest mass density (navy solid line) has been rescaled for a better visualisation.
		}
                The vertical dashed lines represent the known eigenmode frequencies \citep{2013PhRvD..88f4009G}.
		The eigenmode frequencies we obtained are in good agreement with the one in \cite{2013PhRvD..88f4009G, 2014PhRvD..89j4029N}.
		}
	\label{fig:M1_hot_ns_1d_fft}
\end{figure}

{
The neutron star relaxes to its new equilibrium configurations during the first few milliseconds in our simulation, where the surface of the star is neutron-rich ($Y_e \sim 0.1$) with high temperature ($T \sim 10~{\rm MeV}$), which is similar to the case reported in \cite{2014PhRvD..89j4029N}.
The upper panel of figure~\ref{fig:M1_hot_ns_1d_fft} shows the relative variation of the central rest mass density $\rho_c$ in time while the middle panel shows the time evolution of far-field neutrino luminosities $L_\nu$ measured by an observer with fluid at 100 km.
Since the neutrino signals take around 3 ms to reach the extraction point, the luminosities of all neutrinos are zeros before 3 ms.
The luminosities we obtained at the stationary state are at the order of $10^{51}$ erg/s, which are slightly larger than the one reported in \cite{2014PhRvD..89j4029N} (of the order of $10^{50-51}$ erg/s) while much lower than the one reported in \cite{2013PhRvD..88f4009G} (of the order of $10^{52-53}$ erg/s).
In addition, the luminosity of heavy-lepton neutrino $L_{\nu_{x}}$ is found to be highly oscillating.
This is mainly because the opacities and kernels for low density region ($\rho \le 10^{11} {\rm ~ g / cm^{3}}$) are ignored while the interactions of heavy-lepton neutron is sensitive in this region.
}

{
The lower panel of figure~\ref{fig:M1_hot_ns_1d_fft} shows the fast Fourier transform of the central rest mass density, and the luminosities of electron and anti-electron neutrinos with the time window $t \in \left[ 5, 20\right]$ ms.
The fast Fourier transform of the luminosity of heavy-lepton neutrino is not included due to its highly oscillatory nature and the frequency does not correspond to any of the known normal mode oscillations of the neutron star.
To better visualise the result, the amplitude of the fast Fourier transform of the central rest mass density has been rescaled by a factor of a thousand.
The eigenmode frequencies obtained from our simulations agree the one in \cite{2013PhRvD..88f4009G, 2014PhRvD..89j4029N}.
}

Despite the same model, the neutrino luminosities are expected to be different from \cite{2013PhRvD..88f4009G, 2014PhRvD..89j4029N} for two reasons:
(i) the neutrino luminosities are highly sensitive to the neutrino and/or atmosphere treatment adopted in the simulation code \citep{2014PhRvD..89j4029N},
and (ii) the neutrino interactions considered here is not identical to the one in \cite{2013PhRvD..88f4009G, 2014PhRvD..89j4029N}, which could significantly alter the outcome of the neutrino signals.
The test here is to qualitatively compare the results reported in the literature, detailed investigations on {a better low density treatment and} how the neutrino treatment affect the hot neutron star modelling will be left as future work.

\section{\label{sec:conclusions}Conclusions}
We present the new implementation of two-moment based multi-frequency multi-species general-relativistic radiation hydrodynamics module in our code \texttt{Gmunu}.

Our implementation has been tested with several benchmarking tests, which range from special-relativistic to general-relativistic, from optically thick to optically thin and from frequency-integrated to frequency-dependent cases.
These test results demonstrate that our code \texttt{Gmunu} is able to capture the evolution of radiation fields even in the mildly relativistic cases in either optically thin or thick regime.

In addition, we demonstrate that our implicit solvers can robustly solve the largely coupled system, where all the neutrino species at different frequency-bins are coupled altogether, by preforming simulations of a collapsing massive star and a hot neutron star. 
In the core-collapse supernova test, we present the pre-bounce, core-bounce and early post-bounce evolution of a 15 $\rm{M_{\odot}}$ progenitor star.
Also, we show the simulated fair-field neutrino root mean squared energies and luminosities, which are essential to observational astrophysics.
Despite these neutrino quantities are highly sensitive to the implementation of the radiation transport and implicit treatment for neutrino-matter interaction, \texttt{Gmunu} produces consistent result which agrees with other neutrino transport codes.
On the other hand, in the hot neutron star test, our normal mode frequencies are agree with the one reported in the literature.
Moreover, the order of magnitude of the neutrino luminosities extracted from our simulation lie between the one presented in \cite{2013PhRvD..88f4009G, 2014PhRvD..89j4029N} for the same model.
Despite the neutrino treatment and interaction considered are different, our results are qualitatively agree with the literature.

Although our current implementation works properly for the test problems presented, further investigations are needed to improve the module.
In particular, the fluid acceleration terms are not {properly} included in this work.
Including these terms and assessing their impacts on mildly/highly relativistic are essential to neutron star merger simulations.
Besides, solving the entire evolution system including the fluid quantities consistently would lead to more stable and accurate simulations.
Implementing more advance full implicit treatments which allow us to fully solve the system with reasonable computational cost (e.g. \cite{2019ApJS..241....7S, 2021ApJS..253...52L}) is also important for astrophysical applications.
These aspects will be investigated in future work.

\begin{acknowledgments}
We wish to thank Francois Foucart for the helpful discussions on implementations and tests, and detailed comments on the manuscript. 
We also wish to thank David Radice for the helpful discussions on the treatment of frequency-space advection and initial guess of the implicit step, and Tia Martineau for the detailed proofreading and suggestions on the manuscript.
P.C.K.C. acknowledges support from NSF Grant PHY-2020275 (Network for Neutrinos, Nuclear Astrophysics, and Symmetries (N3AS)).
H.H.Y.N. is supported by the ERC Advanced Grant “JETSET: Launching, propagation and emission of relativistic jets from binary mergers and across mass scales” (Grant No. 884631).
The simulations in this work have been performed on the third UNH supercomputer Marvin, also known as Plasma, which is supported by NSF/MRI program under grant number AGS-1919310. 
This work was partially supported by grants from the Research Grants Council of the Hong Kong (Project No. CUHK14306419), the Croucher Innovation Award from the Croucher Fundation Hong Kong and by the Direct Grant for Research from the Research Committee of the Chinese University of Hong Kong.
\end{acknowledgments}

%



\software{
{
The results of this work were produced by utilising \texttt{Gmunu} \citep{2020CQGra..37n5015C, 2021MNRAS.508.2279C, 2022ApJS..261...22C, 2023_leakage}, where the tabulated neutrino interaction were provided with \texttt{Nulib} \citep{2015ApJS..219...24O}.
\texttt{GR1D} \citep{2010CQGra..27k4103O, 2015ApJS..219...24O} was also used to generate one of the reference solutions.
We also modified \texttt{XNS} \citep{2011A&A...528A.101B,XNS1,XNS2,XNS3} to generate the initial data of a hot neutron star.
The data of the simulations were post-processed and visualised with 
\texttt{yt} \citep{2011ApJS..192....9T},
\texttt{NumPy} \citep{harris2020array}, 
\texttt{pandas} \citep{reback2020pandas, mckinney-proc-scipy-2010},
\texttt{SciPy} \citep{2020SciPy-NMeth} and
\texttt{Matplotlib} \citep{2007CSE.....9...90H, thomas_a_caswell_2023_7697899}.
}
}


\appendix

\section{\label{sec:jaco} Jacobian of the monochromatic source terms}
Note that the emission/absorption and elastic scattering source terms $\mathcal{S}^\mu_{\text{E/A}}$ and $\mathcal{S}^\mu_{\text{ES}}$ are monochromatic, i.e. at a given radiation frequency $\nu$, the calculation of the radiation emissivity, absorption and scattering coefficients $\eta \left( \nu \right)$, $\kappa_a \left( \nu \right)$ and $\kappa_s \left( \nu \right)$ do not depend on other radiation frequencies $\nu' \neq \nu $, as shown in equations~\eqref{eq:e_and_a_source} and \eqref{eq:iso_scattering_source}.
The interaction source terms can be largely simplified when only these two source terms are considered.
In this case, the radiation four-force $\mathcal{S}^\mu_{\text{rad}}$ can be reduced to
\begin{equation}
	\begin{aligned}
		\mathcal{S}^\mu_{\text{rad}} = &\mathcal{S}^\mu_{\text{E/A}} + \mathcal{S}^\mu_{\text{ES}} \\
		= &\left( \eta - \kappa_a \mathcal{J} \right) u^\mu - \left( \kappa_a + \kappa_s \right) \mathcal{H}^\mu,
	\end{aligned}
\end{equation}
The corresponding $3+1$ radiation-fluid interaction source terms $\bm{s}_{\text{rad}}$ now becomes
\begin{align}
	&{s_{\text{rad}}}_{\mathcal{E}} = \alpha \psi^6 \sqrt{\bar{\gamma}/\hat{\gamma}} W \Big\{ \eta + \kappa_s \mathcal{J} - \kappa_{as} \left( \mathcal{E} - \mathcal{F}_i v^i \right) \Big\} , \\
	&{s_{\text{rad}}}_{{\mathcal{F}}_i} = \alpha \psi^6 \sqrt{\bar{\gamma}/\hat{\gamma}} \Big\{ 
		  \left( \eta - \kappa_a \mathcal{J} \right) Wv_i - \kappa_{as} \mathcal{H}_i \Big\} , 
\end{align}
where $\kappa_{as} \equiv \kappa_a + \kappa_s$ is the opacity (absorption plus scattering coefficients).
These are the source terms adopted in most of the grey moment codes (e.g. \cite{2022MNRAS.512.1499R}). 
The corresponding Jacobian ${\partial \left[s_{\text{rad}}\right]_i}/{\partial q_j}$, which is needed to calculate the Jacobian in the implicit step discussed in section~\ref{sec:source_terms}, can be evaluated analytically as in \cite{2022MNRAS.512.1499R}.
For completeness, we include the detailed expression of ${\partial \left[s_{\text{stiff}}\right]_i}/{\partial q_j}$, where 
\begin{equation}
	s_{\text{stiff}} \equiv \frac{s_{\text{rad}}}{\psi^6 \sqrt{\bar{\gamma}/\hat{\gamma}}}
\end{equation}
is the undentised source terms.
Below, we denote
\begin{equation}
	\hat{J}_{ab} \equiv \frac{\partial \left[s_{\text{stiff}}\right]_a }{\partial q_b}
\end{equation}
where $a,b = 0, 1, 2, 3$.
Specifically, $\hat{J}_{ab}$ are
\begin{align}
	\hat{J}_{00} =& - W \left( \kappa_{as} - \kappa_s \frac{\partial \mathcal{J}}{\partial \mathcal{E}} \right), \\
	\hat{J}_{0j} =& W \left( \kappa_{as} v^j + \kappa_s \frac{\partial \mathcal{J}}{\partial \mathcal{F}_j} \right), \\
	\hat{J}_{i0} =& - \left( \kappa_{as} \frac{\partial \mathcal{H}_i}{\partial \mathcal{E}} + W \kappa_a \frac{\partial \mathcal{J}}{\partial \mathcal{E}} v_i \right), \\
	\hat{J}_{ij} =& - \left( \kappa_{as} \frac{\partial \mathcal{H}_i}{\partial \mathcal{F}_j} + W \kappa_a \frac{\partial \mathcal{J}}{\partial \mathcal{F}_j} v_i \right).
\end{align}
The corresponding derivatives are
\begin{align}
	&\begin{aligned}
		\frac{\partial \mathcal{J}}{\partial \mathcal{E}} =& W^2 + d_{\text{thin}} \left( v \cdot \hat{f} \right)^2 W^2 \\
				& + d_{\text{thick}} \frac{\left( 3-2 W^2\right)\left(W^2-1\right)}{1+2W^2}
	\end{aligned} , \\
	&\frac{\partial \mathcal{J}}{\partial \mathcal{F}_j} = \mathcal{J}^v_\mathcal{F} v^j + \mathcal{J}^f_\mathcal{F} \hat{f}^j, \\
	&\frac{\partial \mathcal{H}_i}{\partial \mathcal{E}} = \mathcal{H}^v_\mathcal{E} v_i + \mathcal{H}^f_\mathcal{E} \hat{f}_i, \\
	&\begin{aligned}
		\frac{\partial \mathcal{H}_i}{\partial \mathcal{F}_j} =& 
			\mathcal{H}^{\delta}_\mathcal{F} \delta_i^j 
			+ \mathcal{H}^{vv}_\mathcal{F} v_i v^j 
			+ \mathcal{H}^{ff}_\mathcal{F} \hat{f}_i \hat{f}_j \\
			&+ \mathcal{H}^{vf}_\mathcal{F} v_i \hat{f}_j
			+ \mathcal{H}^{fv}_\mathcal{F} \hat{f}_i v_j
	\end{aligned} ,
\end{align}
where we have defined $\hat{f}_i \equiv \mathcal{F}_i / \sqrt{\mathcal{F}^2}$, and
\begin{align}
	\mathcal{J}^v_\mathcal{F} =& 2W^2 \left[ -1 + d_{\text{thin}} \mathcal{E}\frac{v \cdot \hat{f} }{\mathcal{F}} + 2 d_{\text{thick}} \frac{W^2-1}{1+2W^2} \right], \\
	\mathcal{J}^f_\mathcal{F} =& - 2 d_{\text{thin}} W^2 \mathcal{E}\frac{\left(v \cdot \hat{f}\right)^2 }{\mathcal{F}}, \\
	\mathcal{H}^v_\mathcal{E} =& W^3 \left[ -1 - d_{\text{thin}} \left( v \cdot \hat{f} \right)^2 + d_{\text{thick}} \frac{2W^2-3}{1+2W^2} \right], \\
	\mathcal{H}^f_\mathcal{E} =& - d_{\text{thin}} W \left( v \cdot \hat{f} \right) \\
	\mathcal{H}^{\delta}_\mathcal{F} =& W \left[ 1 - d_{\text{thin}} \mathcal{E}\frac{\left(v \cdot \hat{f}\right) }{\mathcal{F}} - d_{\text{thick}} v^2 \right], \\
	\mathcal{H}^{vv}_\mathcal{F} =& 2 W^3 \left[ 1 - d_{\text{thin}} \mathcal{E}\frac{\left(v \cdot \hat{f}\right) }{\mathcal{F}} 
		- d_{\text{thick}} \left( 1 - \frac{ 4W^2 + 1}{2W^2 \left( 2W^2 + 1\right)} \right) \right], \\
	\mathcal{H}^{ff}_\mathcal{F} =& 2 d_{\text{thin}} W \mathcal{E}\frac{\left(v \cdot \hat{f}\right) }{\mathcal{F}}, \\
	\mathcal{H}^{vf}_\mathcal{F} =& - W \mathcal{J}^f_\mathcal{F}, \\
	\mathcal{H}^{fv}_\mathcal{F} =& - d_{\text{thin}} W \frac{ \mathcal{E} }{\mathcal{F}}.
\end{align}

Moreover, the source terms in the evolution equation of electron fraction $Y_e$ is also simplified in this case.
In particular, the contraction of the interaction source and four-velocity term $s_{\rm{rad}}^\mu u_\mu$ in the right-hand-side of equation~\ref{eq:evolution_ye} can now be written as:
\begin{equation}
	s_{\rm{rad}}^\mu u_\mu = \eta - \kappa_a \mathcal{J}.
\end{equation}

\section{\label{sec:decomposed_variables} Decomposition of fluid-frame moments}
Although the fluid-frame moments $\left\{ \mathcal{J}, \mathcal{H}^{\mu} \right\} $ can be computed by contracting the energy momentum tensor $\mathcal{T}^{\mu\nu}$ with the comoving velocities $u^\mu$, it is useful to work out the contraction further, and directly express the fluid-frame moments $\left\{ \mathcal{J}, \mathcal{H}^{\mu} \right\} $ in terms of observer-frame moments $\left\{ \mathcal{E}, \mathcal{F}^{\mu} \right\} $.
As in \cite{spectrecode, 2022MNRAS.512.1499R}, we decompose $\left\{ \mathcal{J}, \mathcal{H}^{\mu}, \mathcal{H}^{\mu}\mathcal{H}_{\mu} \right\} $ as
\begin{align}
	&\mathcal{J} = J_{(1)} + d_{\text{thin}} J_{(2)} + d_{\text{thick}} J_{(3)}, \\
	&\begin{aligned}
		\mathcal{H}_\mu = & - \left( {H_{(1)}}+ d_{\text{thin}} {H_{(2)}}+ d_{\text{thick}} {H_{(3)}}\right) n_\mu \\
				  & - \left( {H_{(4)}}+ d_{\text{thin}} {H_{(5)}}+ d_{\text{thick}} {H_{(6)}}\right) v_\mu \\
				  & - \left( {H_{(7)}}+ d_{\text{thin}} {H_{(8)}}+ d_{\text{thick}} {H_{(9)}}\right) \mathcal{F}_\mu
	\end{aligned} \\
	&\begin{aligned}
		\mathcal{H}^2 = & {\left[H^2\right]}_{(1)} + d_{\text{thin}}{\left[H^2\right]}_{(2)} + d_{\text{thick}}{\left[H^2\right]}_{(3)} \\
		                & + d_{\text{thin}}^2 {\left[H^2\right]}_{(6)} + d_{\text{thick}}^2 {\left[H^2\right]}_{(5)} + d_{\text{thin}} d_{\text{thick}} {\left[H^2\right]}_{(4)},
	\end{aligned} 
\end{align}
where
\begin{align}
	&J_{(1)} = W^2 \left( \mathcal{E} - 2 v^i \mathcal{F}_i \right) \\
	&J_{(2)} = W^2 \mathcal{E} \frac{ \left(v^i \mathcal{F}_i\right)^2 }{\mathcal{F}^j \mathcal{F}_j}  \\
	&J_{(3)} = \frac{ W^2 - 1 }{ 2W^2 + 1} \left[ \left( 3 -2W^2 \right) \mathcal{E} + 4 W^2 v^i \mathcal{F}_i \right] \\
	&H_{(1)} = W \left( J_{(1)} + v^i \mathcal{F}_i - \mathcal{E} \right) \\
	&H_{(2)} = W J_{(2)} \\
	&H_{(3)} = W J_{(3)} \\
	&H_{(4)} = W J_{(1)} \\
	&H_{(5)} = H_{(2)} \\
	&\begin{aligned}
		H_{(6)} = & H_{(3)} + \frac{W}{2W^2+1} \left[ \left( 3-2W^2\right)\mathcal{E} - \left( 1 - 2W^2 \right) v^i \mathcal{F}_i \right]
	\end{aligned} \\
	&H_{(7)} = - W \\
	&H_{(8)} = W \mathcal{E} \frac{ \left(v^i \mathcal{F}_i\right) }{\mathcal{F}^i \mathcal{F}_i}  \\
	&H_{(9)} = W v^i v_i \\
	&\begin{aligned}
		{\left[H^2\right]}_{(1)} = & -{H}_{(1)}^2 +{H}_{(4)}^2 v^2 +{H}_{(7)}^2 \mathcal{F}^2 \\ 
			& + 2 {H}_{(4)}{H}_{(7)} v^i \mathcal{F}_i 
	\end{aligned} \\
	&\begin{aligned}
		{\left[H^2\right]}_{(2)} = & 2 \big( {H}_{(4)}{H}_{(5)} v^2 + {H}_{(7)}{H}_{(8)} \mathcal{F}^2 \\
	       		& + {H}_{(4)}{H}_{(8)} v^i \mathcal{F}_i + {H}_{(7)}{H}_{(5)} v^i \mathcal{F}_i \\
			& - {H}_{(1)}{H}_{(2)} \big) 
	\end{aligned} \\
	&\begin{aligned}
	{\left[H^2\right]}_{(3)} = & 2 \big( {H}_{(4)}{H}_{(6)} v^2 + {H}_{(7)}{H}_{(9)} \mathcal{F}^2 \\
			& + {H}_{(4)}{H}_{(9)} v^i \mathcal{F}_i + {H}_{(7)}{H}_{(6)}  v^i \mathcal{F}_i \\
			& - {H}_{(1)}{H}_{(3)} \big)
	\end{aligned} \\
	&\begin{aligned}
	{\left[H^2\right]}_{(4)} = & 2 \big( {H}_{(5)}{H}_{(6)} v^2 + {H}_{(8)}{H}_{(9)} \mathcal{F}^2 \\
			& + {H}_{(5)}{H}_{(9)} v^i \mathcal{F}_i + {H}_{(8)}{H}_{(6)}  v^i \mathcal{F}_i \\
			& - {H}_{(2)}{H}_{(3)} \big) 
	\end{aligned} \\
	&\begin{aligned}
		{\left[H^2\right]}_{(5)} = & -{H}_{(3)}^2 +{H}_{(6)}^2 v^2 +{H}_{(9)}^2 \mathcal{F}^2 \\
			& + 2 {H}_{(6)}{H}_{(9)} v^i \mathcal{F}_i 
	\end{aligned} \\
	&\begin{aligned}
		{\left[H^2\right]}_{(6)} = & -{H}_{(2)}^2 +{H}_{(5)}^2 v^2 +{H}_{(8)}^2 \mathcal{F}^2 \\
			& + 2 {H}_{(5)}{H}_{(8)} v^i \mathcal{F}_i 
	\end{aligned} 
\end{align}


\bibliography{references}{}
\bibliographystyle{aasjournal}



\end{document}